\documentclass[conference]{IEEEtran}
\usepackage{epsfig,endnotes}
\date{}

\usepackage{epsfig,endnotes}

\newcommand{\PublicRepo}{\url{https://github.com/gsuareztangil/cryptomining-malware}}

\newcommand{\review}[1]{\textcolor{black}{#1}}

\newcommand{\done}[1]{\textcolor{blue}{#1}}

\newcommand{\draft}[1]{{\textcolor{red}{[DRAFT] #1}}}
\newcommand{\todo}[1]{\textcolor{red}{\textbf{TODO:} #1}}

\newif\ifcomment
\commentfalse
\ifcomment
\newcommand{\guillermo}[1]{{\bf \textcolor{blue}{GST: #1}}}
\newcommand{\sergio}[1]{{\bf \textcolor{ForestGreen}{SP: #1}}}
\newcommand{\future}[1]{{\bf \textcolor{brown}{FW: #1}}}
\else
\newcommand{\guillermo}[1]{}
\newcommand{\sergio}[1]{}
\newcommand{\future}[1]{}
\renewcommand{\done}[1]{}
\renewcommand{\draft}[1]{}
\renewcommand{\todo}[1]{}
\fi

\newcommand{\FreebufCampaign}{C\#627}
\newcommand{\PhotominerCampaign}{C\#8}

\usepackage{framed}
\usepackage{subfig}
\usepackage{multirow}
\usepackage{paralist}
\usepackage{comment}
\usepackage{booktabs}

\usepackage{fontawesome}
\usepackage{rotating}

\newcommand{\invertedFaGlobe}{\begin{sideways}%
      \begin{sideways}\faGlobe\end{sideways}\end{sideways}}

\newcommand{\malware}{\includegraphics[width=.4cm, trim=0 0 0 -75]{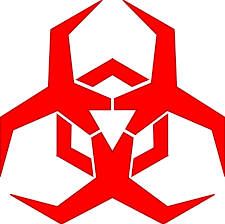}}

\usepackage{color}
\usepackage{tabularx}
\usepackage{pifont}
\usepackage{siunitx}
\usepackage{url}

\usepackage{ifthen}
\newif\ifshort
\shortfalse %
\ifshort
	\newcommand{\isShort}{true}
\else
	\newcommand{\isShort}{false}
\fi
\newcommand{\shortVer}[1]{\ifthenelse{\equal{\isShort}{true}}{{#1}}{}}
\newcommand{\longVer}[1]{\ifthenelse{\equal{\isShort}{false}}{{#1}}{}}

\IEEEoverridecommandlockouts

\newcommand{\descr}[1]{\smallskip\noindent\textbf{#1}}
\newcommand{\descrit}[1]{\vspace{0.05cm}\noindent{#1}}

\begin{document}

\title{A First Look at the Crypto-Mining Malware Ecosystem: A Decade of Unrestricted Wealth\textsuperscript{*}\thanks{\textsuperscript{*}A shorter version of this paper appears in the 2019 ACM Internet Measurement Conference (IMC). This is the full version.}} %

\thispagestyle{plain}
\pagestyle{plain}

\author{
{\rm Sergio Pastrana}\\
Universidad Carlos III de Madrid\\
\url{spastran@inf.uc3m.es}
\and
{\rm Guillermo Suarez-Tangil}\\
King's College London\\
\url{guillermo.suarez-tangil@kcl.ac.uk}
} %

\maketitle

\begin{abstract}
Illicit crypto-mining leverages resources stolen from victims to mine cryptocurrencies on behalf of criminals. While recent works have analyzed one side of this threat, i.e.: web-browser cryptojacking, only commercial reports have partially covered binary-based crypto-mining malware. 

In this paper, we conduct the largest measurement of crypto-mining malware to date, analyzing approximately 4.5 million malware samples (1.2 million malicious miners), over a period of twelve years from 2007 to 2019. 
Our analysis pipeline applies both static and dynamic analysis to extract information from the samples, such as wallet identifiers and mining pools. 
Together with OSINT data, this information is used to group samples into campaigns.  
We then analyze publicly-available payments sent to the wallets from mining-pools as a reward for mining, and estimate profits for the different campaigns. All this together is is done in a fully automated fashion, which enables us to leverage measurement-based findings of illicit crypto-mining at scale.

Our profit analysis reveals campaigns with multi-million earnings, associating over 4.4\% of Monero with illicit mining. 
We analyze the infrastructure related with the different campaigns, showing that a high proportion of this ecosystem is supported by underground economies such as Pay-Per-Install services. We also uncover novel techniques that allow criminals to run successful campaigns. 


\end{abstract}

\section{Introduction}

Mining is a key component responsible for the wealth of Blockchain-based cryptocurrencies. This process requires a network of interconnected miners to solve a complex mathematical problem in order to link blocks and maintain the integrity of the transactions. In exchange, miners receive an amount of the mined cryptocurrency as a reward.

The high value of cryptocurrencies has attracted a large number of malicious actors that use hijacked resources to mine these currencies. 
The illicit crypto-mining threat has grown considerably over the recent years
~\cite{McAfee18},  %
and it is considered one of the top-most cybersecurity threats, even surpassing ransomware according to recent reports~\cite{Webroot}.

Illicit crypto-mining is typically conducted using either one of these two modes: 
\begin{inparaenum}[(i)]
\item by using {\tt browser}-based crypto-mining programs (dubbed cryptojacking~\cite{lastline2018}), where the mining process is run in scripts (typically JavaScript) embedded in web content; or
\item by using {\tt binary}-based crypto-mining malware, where the mining process is embedded in the payload of a malware running in infected machines that are connected to the Internet.
\end{inparaenum}
In both cases, by using hundreds of hijacked machines, perpetrators can obtain a hash-rate similar to medium-sized mining farms. Each mode has different characterizing features and unique challenges, specially when it comes to devising effective countermeasures. For example, in browser-based cryptojacking the damage ceases when the victim stops browsing the site. Also, users can reduce the threat by restricting the use of JavaScript. 
Meanwhile, crypto-mining malware entails classical malware-related challenges, such as persistence and obfuscation. Also, since mining increases the CPU load, thus reducing the computer's performance, it might be noticed by end-users. Thus, we observe a new paradigm aimed at evading user- rather than AntiVirus-detection using techniques such as \textit{idle mining} (mining only when the CPU is idle) or reducing CPU consumption when monitoring tools (e.g., Task Manager) are running. %
\longVer{For readers unfamiliar with the topic, we refer to the {\it Background} section in~\S\ref{sec:background} for an introduction to cryptocurrency mining and its threats.}

\descr{Motivation.}  
While illicit crypto-mining has been less notorious than other threats such as ransomware, it poses nonetheless an important threat to users and organizations; and its presence is an indicator of weaknesses in security practices that must be addressed~\cite{cyberthreatalliance2018}. 
First, the profits generated by their miners introduce massive incomes to cyber-criminals. 
These incomes fuel the underground economy and gear other cyber-criminal activities~\cite{Thomas15framingdependencies}.
Second, this threat causes important economical loses to their victims. 
By draining the CPU-usage, corporations see how their electricity bills increase and how their hardware rapidly deteriorates~\cite{Webroot,Tahir17}. 
Finally, this indirectly causes a non-negligible environmental footprint~\cite{kirat2014barecloud}.  
Due to these concerns,
browser-based crypto-mining has been widely studied recently, both analyzing it as a crime~\cite{Konoth18,Hong18,Saad18,Kharraz19} and as an alternative business model to monetize web content~\cite{Ruth18,Papadopoulos18,Musch18}. However, the literature lacks of a systematic approach to %
measure the binary-based mining threat at scale.%

The first and only seminal work putting this threat in
perspective is from 2014~\cite{Huang2014}. 
Authors analyzed 2K malware samples mining Bitcoin and their methodology relied on the analysis of public transactions. 
However, there has been a significant increase in the number of malware samples monetizing this threat since 2014~\cite{McAfee18,minersPaloAltoNetworks}.
Also, criminals' attention has shifted to other cryptocurrencies, mainly motivated by: i)  the proliferation of ASIC mining, which uses dedicated hardware and renders the use of desktop computers no longer profitable for mining bitcoins, and ii) the development of protocols that provide transaction anonymity (such as those used in Zcash or Monero). 
Anonymous currencies are used by criminals to thwart traceability and they are on demand in underground markets.
Commercial reports, in the form of blog posts~\cite{minersPaloAltoNetworks,lastline2018} or white papers~\cite{cyberthreatalliance2018} provide a further, but limited, view of the magnitude of the problem and the landscape. %
Security firms have analyzed isolated cases of decontextualized mining operations~\cite{Smominru,Adylkuzz}. 
However, these studies are limited by the simplicity of the analysis. %

In this paper, we aim to bridge these gaps by addressing the following research questions:
\begin{inparaenum}[(1)]
  \item What are the preferred cryptocurrencies mined by criminals?
  \item How many actors are involved in this ecosystem and what are their profits?
  \item What is the level of sophistication used in different campaigns and how does this affect the earnings? 
  \item What is the role of underground markets and what are the tools and techniques adopted from them?
  \item How can we improve current countermeasures and intervention approaches? %
\end{inparaenum}
Due to potential {\em ethical concerns} arisen from this work (see Appendix), we obtained approval from our REB office.

\longVer{\descr{Novelty.}}
\shortVer{\descr{Contributions}.}
Our work focuses on crypto-mining malware\shortVer{as key novelty}. 
By looking at a wide-range of underground communities, where knowledge and tools are shared,
we have observed increased interest in this malware.
This suggests that cybercrime commoditization plays a key role in the wealth of illicit crypto-mining.
We design a measurement pipeline to %
automatically analyze malware samples observed in the wild and to extract information required to identify the miners and pools, using both dynamic and static analysis.  %
Then, we build a graph-based system that aggregates related samples into campaigns based on a series of heuristics. 
The system is designed to distinguish campaigns using third-party infrastructure such as Pay-Per-Install (PPI) services or binary obfuscators. %
This allows to analyze to what extent this threat is sustained by different underground markets~\cite{Thomas15framingdependencies}. 
Our analysis system %
enables the research community to leverage crypto-mining measurements at scale.
\shortVer{To the best of our knowledge, this paper presents the largest systematic study of malicious binary-based crypto-mining, providing a reliable lower bound of the earnings made by this criminal industry.}

\longVer{
\descr{Findings.}
Among others, our main findings include:
\smallskip
\begin{asparaenum}
\item Monero (XMR) is by far the most popular cryptocurrency among cyber-criminals in underground economies (\S\ref{finding:underground}). %
Considering only crypto-mining malware, our profit analysis shows that criminals have mined over 4.37\% of the circulating XMR. 
Although this depends on when criminals cash-out their earnings, we estimate that the total revenue accounts for nearly 58M USD. These criminal earnings should be added to estimations from parallel work focused on browser-based cryptojacking %
(\S\ref{sect:related-work}).

\item {Campaigns that use third-party infrastructure (typically rented in underground marketplaces) are more successful. However, this is not always the case. Some of }
the most profitable campaigns rely on complex infrastructure that also uses general-purpose botnets to run mining operations {without using third-party infrastructure}. 
Here, we discover novel malware campaigns that are previously unknown to the community (e.g., the code-named {\tt Freebuf} or the {\tt USA-138} campaign presented as case studies in \S\ref{sec:case-studies}\shortVer{ and in the extended version of this paper~\cite{extended} respectively}). {Moreover, only some criminals keep their infrastructure updated, for example when they are banned in mining pools or when the mining software needs to be updated due to changes in the mining algorithm}. 

\item Campaigns use simple mechanisms to evade detection, {like using domain aliases to contact mining pools (which prevents simple blacklisting approaches), or \textit{idle mining}}. %

\item There are other criminals running successful campaigns with minimal infrastructure. A common yet effective approach is to
use legitimate infrastructure such as Dropbox or GitHub to host the droppers, and stock mining tools such as {\tt claymore} and {\tt xmrig} to do the actual mining. 
We also show what are the most popular Monero mining pools ({\tt crypto-pool}, {\tt dwarfpool} and {\tt minexmr}) among criminals and discuss the role of these and other pools when devising countermeasures. 
\end{asparaenum}
}

\shortVer{In summary, the main contributions of this paper are:} \longVer{\descr{Contributions.} To the best of our knowledge, this paper presents the largest systematic study of malicious binary-based crypto-mining.
Our main contributions are: }
\begin{compactenum}
\item {We analyze and describe the role of underground communities for the proliferation of the illicit crypto-mining business~(\S\ref{sec:underground}).} 
\item We present a system that uses both static and dynamic analysis to extract relevant mining-related information from crypto-mining malware, such as wallet addresses and pool domains~(\S\ref{sec:methodology}). 
Our system uses different techniques to aggregate related samples into larger campaigns represented as a graph that is then mined for further analysis. {Additionally, we feed the system with information gathered from various Open-Source Intelligence (OSINT) repositories to further classify and analyze the campaigns.} 
\item We present a {longitudinal} study of the crypto-mining malware threat using data spanning over more than a decade~(\S\ref{sec:analysis} and~\S\ref{sec:case-studies}). Then, by focusing on Monero, we rely on information gathered from mining pools to measure the earnings gained by each campaign. We also analyze the infrastructure used by criminals and extract the attribution to stock mining software. 
\item We propose a number of countermeasures, and discuss the efficacy of existing ones together with the open challenges~(\S\ref{sec:discussion}). Then, we %
contextualize the most important findings of our study with respect to relevant works in the area~(\S\ref{sect:related-work}). 
\end{compactenum}
Finally, to foster research in the area, we release our dataset in our online repository.\footnote{\PublicRepo}  
We encourage readers to visit this repository\shortVer{ and in the extended version of this paper~\cite{extended}} as it provides a wider presentation of the measurements left out of this paper due to space constraints.
\section{Background}
\label{sec:background}

In this section, we first provide an overview of the cryptocurrency mining process. Then %
we describe the underground economy supporting the illicit crypto-mining threat. 

\shortVer{
\descr{Cryptocurrency Mining.}
Cryptocurrencies are digital assets that can be exchanged in online transactions. These transactions are grouped into blocks and added to a distributed database known as the blockchain. Each block is linked to its previous block, and the addition of new blocks is done by voluntary miners. In order to provide integrity, miners compute a cryptographic hash of the block together with the solution to a complex mathematical puzzle known as `Proof-of-Work' (PoW). As a reward, miners receive a certain amount of the cryptocurrency. %

The increased value of cryptocurrencies such as Bitcoin or Ethereum has lead to the growth of mining farms using specialized hardware known as ASICS, which makes end-user machines useless for mining. However, in 2014 a new PoW known as Cryptonote required not only CPU power but also memory, turning ASIC-based mining inefficient. Additionally, the PoW algorithm changes periodically, thus discouraging ASIC development (which is optimized for specific algorithms)~\cite{forkingASIC}. This allows individuals to mine with their end-user machines.  Examples of cryptocurrencies using Cryptonote as PoW are: Monero and Bytecoin.

When mining a block, only the first one solving all PoWs gets the reward.
Thus, mining has become a race which highly depends on the hashrate (i.e., number of hashes computed per second) of a miner. A higher hashrate increases the probability of mining a block. A common approach is to mine through public mining pools, which can be viewed as partnership between workers where the complexity of the mining challenge is distributed among the partners. Each partner contributes with a given hashrate aimed at solving the puzzle, and when the pool successfully mines a block, the reward is divided proportionally to the workers' hashrate. To receive rewards, the workers are identified in the pool. Some pools use proprietary site-keys, like CoinHive, which was the major provider of browser-based mining services until March 2019. Other pools use emails or wallet addresses. The communication between each miner and the pool is done by using the Stratum protocol. This is a clear-text protocol that uses JSON-RPC format to transmit information, including authentication tokens, block puzzles and mined hashes. We refer to the work by Recabarren and Carbunar for a complete analysis of this protocol~\cite{Recabarren17}.  

\textit{Illicit crypto-mining} refers to mining carried out by criminals using resources stolen from their victims. This threat appeared with Bitcoin in 2009~\cite{Huang2014}, but it has increased notably since 2014 due to the inception of PoW algorithms resistant to ASIC-based mining. Illicit crypto-mining is performed in the device of a victim by either embedding a payload in web resources (scripts) executed by browsers~\cite{Hong18,Konoth18}, or distributing the payload in the form of malware.
}

\longVer{
\descr{Cryptocurrency Mining.}
Cryptocurrencies are a type of digital assets that can be exchanged in online transactions. These transactions are grouped into blocks and added to a distributed database known as the blockchain. Each block is linked to its previous block. Addition of new blocks to the blockchain is done by voluntary miners. These must compute a cryptographic hash of the block, which includes complex mathematical puzzle known as `Proof-of-Work' (PoW). As a reward, miners receive a certain amount of the currency. The mining process maintains the integrity of the blockchain and it is at the core of all cryptocurrencies.

The increased value of cryptocurrencies such as Bitcoin or Ethereum leads to the growth of mining farms using specialized hardware known as ASICS. Thus, mining these cryptocurrencies using end-user machines such as laptops or desktop computers was useless. However, in 2014 a new PoW known as Cryptonote required not only CPU power but also memory, turning ASIC-based mining inefficient and thus gaining again the attention of individuals willing to mine with their home machines. Additionally, the mining algorithm changes periodically, thus discouraging ASIC development (which is optimized for specific algorithms)~\cite{forkingASIC}. Examples of cryptocurrencies using Cryptonote as PoW are: Monero (XMR) and Bytecoin (BCN).

When a new block is added to the blockchain, only the first miner being able to mine the block will get the reward. This turns the mining process into a race where speed is the hashrate of a miner. The higher the hashrate, the higher the probability of mining a block and thus getting the reward. Accordingly, mining is typically done using public mining pools, which can be viewed as partnership services between various workers where the complexity of the mining challenge is distributed among the partners. Each partner contributes with a given hashrate aimed at solving the puzzle, and when the pool successfully mines a block, the reward is divided among the partners proportionally to their hashrate.  In order to get this reward, workers must provide some form of identification. This can be proprietary site-keys, like in the case of CoinHive (the major provider of browser-based mining services), emails or wallet addresses. The communication between each miner and the pool is done by using Stratum, which is a de-facto TCP based protocol evolved from the \textit{getwork} protocol~\cite{Recabarren17}. 

\descr{Crypto-mining malware and cryptojacking.}
Crypto-cu\-rren\-cy mining is a rather easy monetization technique using hardware resources. However, it requires an investment in equipment and also entails a cost in terms of energy. In illicit cryptomining, criminals make use of their victims' computing resources to mine cryptocurrencies on their behalf. This threat exists since the creation of Bitcoin in 2009, but it has increased since 2014 due to the inception of Cryptonote and other PoW algorithms resistant to ASIC-based mining. 

Illicit crypto-mining is performed by using two techniques: browser-based or binary-based mining. In the former, the mining payload is embedded in web resources which are executed by client browsers without the explicit consent of the users~\cite{Hong18,Konoth18}. In the latter, the payload is distributed in the form of malware.
}

\descr{The Underground Economy.}
\label{sec:underground}
Underground markets play a key role in the business of malicious crypto-mining. Users with few technical skills can easily acquire services and tools to set up their own mining campaign. Forums are also used for sharing knowledge. 
To put our study in context, we have analyzed a dataset of posts collected from various underground forums~\cite{Pastrana18a}, looking for conversations related to crypto-mining. 
We observe that crypto-mining malware can be easily purchased online, for a few dollars (e.g., the average cost for an encrypted miner for Monero is 35\$).
In particular, we have seen an online service which allow to create customized binaries (e.g., for a particular cryptocurrency and/or a given pool) to mine cryptonote based currencies, for \$13.\footnote{For ethical reasons we do not disclose the URL of this service} It provides several stealthy-related techniques such as idle mining or the use of execution-stalling code~\cite{kolbitsch2011power} targeted to certain conditions (e.g., when the Task Manager is running). 
Other providers opt to share their miners for free, in exchange for a donation: \textit{``Miner is free, we charge a fee of 2\% to cover the time coding.''} 
\longVer{Figure~\ref{fig:evolution-threads-forums}}\shortVer{The extended version of this paper~\cite{extended}} shows a longitudinal analysis of posts related to crypto-mining in these forums. 
Here, we show that Monero is the most prevalent currency nowadays. 

\longVer{
\begin{figure}[h]
    \centering
    \includegraphics[width=.99\columnwidth]{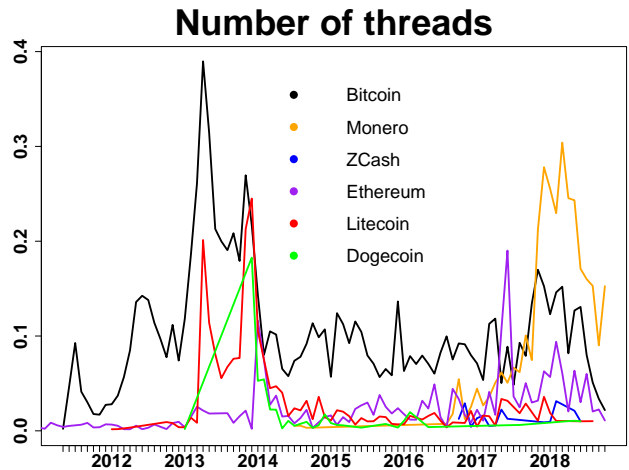}
    \caption{Evolution of the number of threads from underground forums related to mining of different cryptocurrencies.}
    \label{fig:evolution-threads-forums}
\end{figure}
}

We also observed that a common topic of conversation concerns (i) ``friendly'' pools, i.e.: pools that do not generally ban users displaying botnet-like behaviors, or (ii) how to remain undetected otherwise. 
For instance, users claim that a good trade-off between profitable hash-rates and a long-lasting mining strategy is using botnets with less than 2K bots. 
For bigger botnets, many discussions and tutorials explain how to configure proxies and provide advice on how to reduce the risk of being exposed:  \textit{``The best option is to use a proxy and you can use any pool. 
Contact me for PM, I am willing to help''}. 
Also, we found various conversations with users looking for partners and offering custom (private) mining pools: \textit{``In my pool there is no ban by multiple connections.''}

Finally, we note that it is also possible to purchase {\it all-you-need} packages, including tools and services, with a guarantee period and maintenance (e.g., re-obfuscation when the miner is detected, or updates to new versions). 
\longVer
{For the curious reader, \longVer{Figure~\ref{fig:botnet-undergound-forums}}
\shortVer{the extended version of this paper~\cite{extended}} shows a flyer posted in one of the underground markets, which offers a full Monero botnet. 
}

\longVer{
\begin{figure}[th!]
    \centering
    \includegraphics[width=\columnwidth]{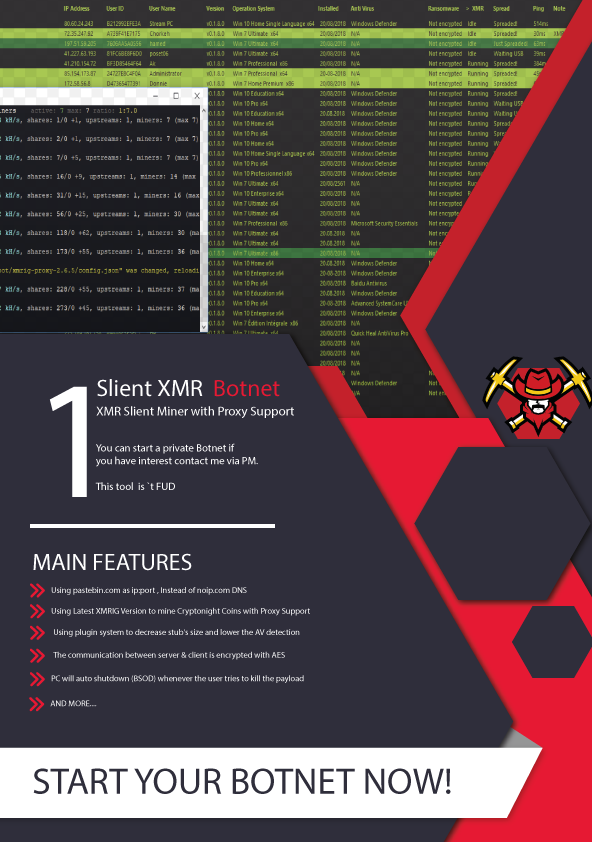}
    \caption{Crypto-mining package offered in an underground forum, including botnet setup, XMR miner and proxy. Permanent link: {https://perma.cc/4FN8-B98M}.}
    \label{fig:botnet-undergound-forums}
\end{figure}
}

\descr{Take-Away:} {The support offered by underground communities to criminals explains the sharp growth on the amount of malware monetizing their victims. This motivates the need for a longitudinal measurement of this threat. We show that Monero is currently the most discussed crypto-mining coin by underground forum users.}
\section{Measurement Methodology}
\label{sec:methodology}
\begin{figure*}
    \centering
    \includegraphics[width=.9\textwidth, trim=0 140 0 0, clip]{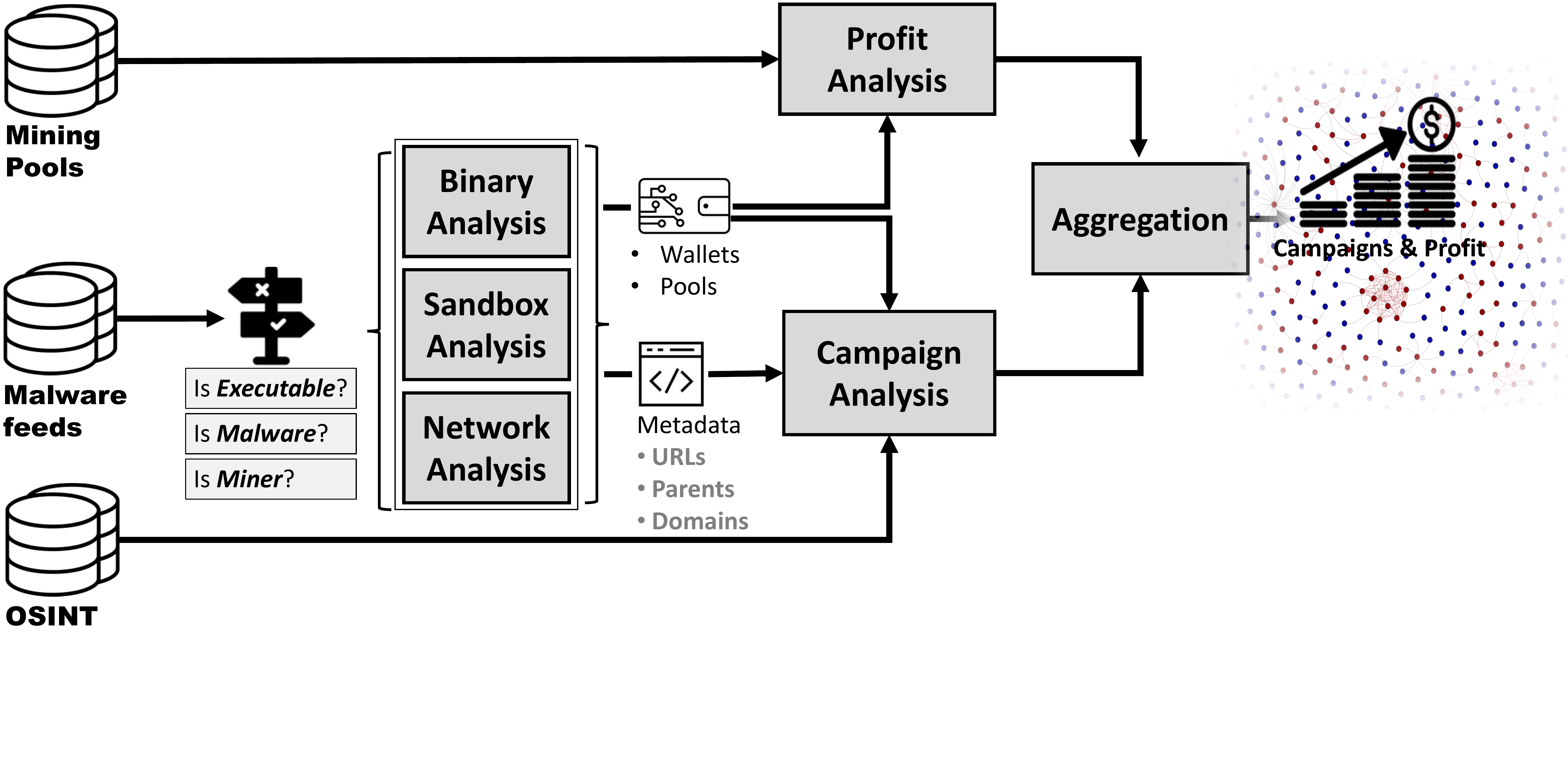}
    \caption{Overview of our processing pipeline and measurement methodology.}
    \label{fig:overview}
\end{figure*}

A general overview of the measurement methodology is
presented in Figure~\ref{fig:overview}. 
For the sample collection, we query both public and private repositories of malware and different intelligence feeds as described in~\S\ref{sec:methodology:data}. 
We make a number of sanity checks for each sample to ensure that we only feed crypto-mining malware to our pipeline (see~\S\ref{sec:methodology:sanity}). 
We also collect OSINT related to running botnets and relevant Indicators of Compromise (IoCs) observed in malware samples. 

A key phase in our pipeline is to analyze relevant samples both statically and dynamically as described in~\S\ref{sec:methodology:analysis} 
({\em Binary}, {\em Sandbox}, and {\em Network Analysis} in Figure~\ref{fig:overview}). 
The goal of this multi-step phase is to extract the following information from each miner:
\begin{inparaenum}[(i)]
  \item the pool or address which the crypto-mining malware connects to for mining, and the identifier used to authenticate themselves into this pool,
  \item the addresses of the e-wallets where mined cryptocoins are paid to\footnote{The terms {\it address} and {\it wallet} are used interchangeably in the literature.}, which in most cases coincide with the identifier, 
  \item URLs where the malware connects to or is seen at, and 
  \item other metadata obtained from intelligence feeds such as when the sample was first seen or related samples (e.g., dropped binaries).
\end{inparaenum}

The next step is to analyze the mining pools that the miners work with. 
We decouple connections made to proxies (that in turn connect to pools) from the connections to the actual pools. %
We then look at the profit reported by each of the wallets in a pool. 
These two steps are described in~\S\ref{sec:methodology:proxy-profit} and are referred to as {\em {Profit Analysis}} in Figure~\ref{fig:overview}.

Finally, we aggregate related samples into campaigns and analyze them separately in the {\em Aggregation} step as described in~\S\ref{sec:methodology:aggregation}. %
For this, we create a graph which interconnects crypto-mining malware that: 
\begin{inparaenum}[(i)]
  \item share a common execution ancestor (i.e., dropper) or are packed together, 
  \item accumulate their earnings into the same wallet, 
  \item share common infrastructure (e.g., proxies or hosting servers), or
  \item relate to the same IoC related with mining campaigns --- gathered both from OSINT reports and our own investigation. 
\end{inparaenum}
We then enrich every interconnected sub-graph (campaign) to include details about related infrastructure used in each campaign (i.e., stock mining software or Pay-Per-Install services). 

\subsection{Data Gathering}
\label{sec:methodology:data}

We collect malware samples, metadata and OSINT information, and known mining tools from various sources.

\descr{Malware}.
We rely on public and private feeds from:

\begin{compactenum}
\item {\tt \textbf{Virus Share}}. This is an online community that shares torrents to malicious binaries. 
We use it to gather our initial dataset of raw binaries. 
\item {\tt \textbf{Virus Total}}. This is an online service containing both publicly and privately available services to the security community through an API (Application Programming Interface). 
This service is a subsidiary of Google that runs multiple AntiVirus (AV) engines and offers an unbiased access to resulting reports. 
We use the private API to download malware binaries. 
We then query the public API to obtain \longVer{relevant} threat intelligence (which we refer to as metadata\longVer{ in the paper}). 
In particular, we collect metadata from \longVer{the samples obtained through queries to Virus Total, but also from} all the samples obtained through \longVer{other}\shortVer{all} sources listed in this section. 
\item {\tt \textbf{Hybrid Analysis}}. This is an online community providing malicious binaries and 
threat intelligence obtained from static and dynamic analysis of the samples. 
Thus, we use this community to fetch readily available intelligence when possible.
\item {\tt \textbf{Other Sources}}. We have developed a crawler to fetch samples from a variety of online communities such as {malc0de.com} or {vxvault.net}. 
Our pipeline also aggregates malware feeds from cybersecurity companies. 
For the purpose of this paper, we have received feeds from {Palo Alto Networks}\longVer{ with miners known to them}.  
\end{compactenum}
{Refer to\shortVer{~\cite{extended}} \longVer{Appendix~\ref{sec:appendix-metadata}} for details about the dataset overlaps.}

\descr{Metadata}.
When available, we primarily rely on the following metadata to put our study in context:
\begin{inparaenum}[(i)]
  \item the first time the sample was seen in the wild,  
  \item the {URLs} where the sample was seen,
  \item the list of parents that are known to have dropped the binary under analysis, and
  \item the list of contacted domains.
\end{inparaenum}

\descr{Stock Mining Software}.
We also collect binaries from {\em known mining frameworks}, such as {\tt xmrig}\footnote{\url{https://github.com/xmrig/xmrig}} or {\tt xmr-stak}\footnote{\url{https://github.com/fireice-uk/xmr-stak}}, that are hosted in various public repositories. 
While these binaries are not badware {\it per se}, their usage is deemed malicious when run by malware. 
Our assumption is that the usage of proprietary software to mine is not the norm. 
Anecdotal evidence observed during the course of a preliminary investigation has shown that miscreants rely on legitimate --- open-source --- mining tools. %
The modus operandi of the malware is to fetch one of these tools (i.e., acting as a dropper) and run it in infected machines. Mining is configured with the wallet of the miscreant, where the rewards are paid by the network. 
One of our goals then is to understand if this assumption holds true and how many campaigns are using stock mining software illicitly. %

\descr{Summary}.
Our data collection registers over 4.5 million samples (see \S\ref{sec:analysis:dataset} for a breakdown), which have been active between early 2007 and early 2019. %
This includes about 1K versions of known mining tools from 13 different frameworks. 
Our initial data contains a wide-range of samples, many of which are irrelevant to this study (e.g., web-based cryptojackers). 
Thus, we next describe the rules we use to consolidate the dataset where we report our findings with.

\subsection{Sanity Checks}
\label{sec:methodology:sanity}

One important aspect when systematizing the analysis of malware is properly curating the dataset~\cite{lever2017lustrum}.
We perform the following sanity checks for each sample\longVer{ processed}: 
\begin{inparaenum}[(i)]
  \item {\em is it malware?}
  \item {\em is it a miner?}, and
  \item {\em is it an executable sample?}
\end{inparaenum}

First, we rely on Virus Total reports to learn if a sample is malware. 
Virus Total have been shown to perform remarkably well when providing malware feeds according to a recent comparative analysis of Threat Intelligence~\cite{liUsenix2019osint}. 
In particular, Virus Total was able to detect 99.94\% of the threats over one of the largest non-targeted\footnote{Meaning that they target malware threats to generic platforms. 
Other targeted malware aggregators focus on threats that specifically target platforms like Facebook and ``that are not as relevant to most Virus Total users''~\cite{liUsenix2019osint}.} malware aggregators.
We assume that a sample is malware if at least 10 AV vendors flag the sample as malicious. 
While this is a common practice in other works in the area~\cite{Lindorfer14,Lindorfer15,Kantchelian15}, we acknowledge that having a solid ground-truth is essential (see discussion in {\S\ref{sec:discussion:ground-truth}}). 
Thus, we use a white-list with the hashes of known mining tools,  to ensure that they are not considered as malware samples in our study. This white-list is compiled from binaries collected from various online open-source repositories. 
\review{A wallet extracted from a malware sample is considered an `illicit' wallet throughout our dataset. 
Therefore, we exceptionally keep samples with less than 10 AV positives when it contains an illicit wallet.} %

Second, we assume a malware is a crypto-mining tool when there are IoCs that reveal \review{activity related to mining, such as connections made through the Stratum protocol or DNS resolutions for web mining pools. To this end, }
we apply publicly available YARA rules\footnote{\url{https://github.com/Yara-Rules}} to our samples. 
Additionally, we \review{compare OSINT information related to known mining campaigns} with IoCs extracted from the samples (e.g., file hashes or network data). 
We also use advanced queries from Virus Total and Hybrid Analysis to look for malware that meet the following criteria:
\begin{inparaenum}[(i)]
  \item samples that contact domains of known mining pools,
  \item communicate through the {Stratum} protocol, and
  \item are labeled as ``Miner'' (or related variants) by more than 10 AVs.
\end{inparaenum}

Finally, to understand whether a malware is executable, we rely the magic number from its header, and consider only those related to executables like PE, ELF or JARs. 
\S\ref{sec:discussion:ground-truth} provides discussion of the limitations behind these assumptions. 

\subsection{Extraction of Pools and Wallets}
\label{sec:methodology:analysis}

With our dataset of crypto-mining malware, we rely on: %
\begin{compactenum}
    \item {\it Static Analysis}: we perform binary inspection to extract evidences of mining activity embedded \textit{into the binary}. 
    \item {\it Dynamic Analysis}: we then use environmental information obtained from the \textit{execution of the binary} in a sandbox. Specifically, we obtain the network traffic, the dropped files, processes opened, and command line parameters passed to the binaries. \review{When available, we rely on reports provided by \textit{Virus Total} and \textit{Hybrid Analysis} through their API service.}
\end{compactenum}

In some cases we are able to find identifiers (e.g., wallets or emails) and pool names using static analysis. 
In other cases, we rely on dynamic analysis to extract these identifiers from the network activity or the command line processes.
In both cases, we process the output of these two analyses using heuristics and regular expressions to extract the following information:

\descr{Cryptocurrency wallets}. 
Miners connect to the pools using the Stratum protocol~\cite{Recabarren17}. 
Upon connection to the pool, they send a request-for-work packet with the identifier of the miner in a `login' parameter. 
This identifier can be extracted from the command line options passed to the mining tool or directly from the network traffic. We also process the type of wallet to understand the cryptocurrency (e.g.: Monero, Bitcoin or Ethereum) the malware is intending to mine. 

\descr{Mining pools}. 
We collect additional information such as domains and IPs of mining pools and proxies. Similarly to wallet addresses, this information is typically extracted from either the command line of the process invoking the mining tool or from the network traffic.
Typically, miners connect to a known pool.\footnote{We consider known pools as those listed in public sources, e.g.: \url{http://moneropools.com/} or \url{http://www.blockchain.com/pools}.} 
In some cases, the miner either uses a proxy or mines against a private/unknown pool.\footnote{{While the use of private pools is encouraged in certain underground communities, we have observed few samples using private pools.}} %
We consider that a miner is using a proxy if we record mining activity for the corresponding wallet in a known pool (see~\S\ref{sec:methodology:proxy-profit}).

\subsection{Collecting Mining Activity}
\label{sec:methodology:proxy-profit}

One of the main challenges when measuring the impact of the malicious crypto-mining campaigns is the difficulty to accurately estimate the profits. 
In \longVer{the case of} browser-based cryptojacking, recent works use estimations of the number of visitors per hour for similar websites and the average hashrate of a single visitor (victim)~\cite{Hong18,Konoth18,Eskandari18}. 
This is highly inaccurate as evidenced by the variances reported by concurrent related works (see~\S\ref{sect:related-work}). 
In the case of crypto-mining malware, the actual wallet which the mining reward is paid to can be extracted. %
We leverage public information obtained from mining pools (which include total reward paid to wallets) to get a more approximate estimation of the profits. 

For all the extracted wallets, we queried the most prevalent mining pools to collect activity associated with these wallets. While the amount of information offered by each pool varies, it always contains the timestamp of the last share, the current (last) hashrate and the total amount of currency paid to the wallet. Additionally, some pools also provide the historic hashrate of the wallet and the list of payments done to the wallet (including timestamp and amount). 
While the total paid is always available, some pools only provide payment data for the last period (e.g. a week or month). Since we are interested on studying how the payments evolve across time, we use public APIs to collect this information periodically for a period of 10 months (July'18-April'19). 
As a single wallet can use more than one mining pool, we queried all the wallets against all the pools.
Then, to estimate profits, we aggregate all the payments sent by the pools to the wallets. 
In general, we report payments using XMR. 
To ease readability we also report the equivalent in US dollars (USD). 
However, we note that we do not have information about when the criminals have cashed-out their earnings (if ever). 
Thus, it is hard to extract an exact figure in USD (and other currencies) due to the fluctuations on the value of Monero.
To approximate this value, we dynamically extract the exchange rate between XMR and USD of the date when the payments were made, if available. 
We use the average exchange rate of 54 USD/XMR in cases where historical payments are unavailable.  
\subsection{Campaign Analysis}
\label{sec:methodology:aggregation}

Two major limitations in related works are: i) the simplicity in which they analyze related mining campaigns, and ii) the inability to study anonymous cryptocurrencies such as Monero (as discussed in \S\ref{tab:related_work}).
Thus, \longVer{in this work,} we aggregate samples into campaigns following a novel methodology that leverages various characterizing features observed in the wild. %
\longVer{We emphasize that the methodology we use to aggregate samples into campaigns is novel. 
We also note that our methodology admits a wide range of features.}

\descr{Spreading Infrastructure}. 
We distinguish two types of infrastructure used to spread the malware: one that can be owned and another one that belongs to a third-party and can be rented (e.g., botnets that are monetized as PPI services and that are used for mining). 
When available, we link samples to known botnets by querying OSINT information with IoCs extracted from the samples. 
We refrain from using these botnets to aggregate samples as we detail later. 
However, we use them to enrich the information of the campaigns in a post-aggregation phase. 
This way, we can draw conclusions about the number of campaigns using \textit{known} third-party infrastructure.
However, since we rely on public intelligence feeds, a limitation of this approach is that samples using \textit{unknown} third-party infrastructure (e.g., offered in underground markets) might be aggregated together in a single campaign. 
In these cases, we can guarantee that the campaign runs on top of the same infrastructure. 
This is relevant to law enforcement agencies when devising take-downs strategies. 
Thus, our analysis considers campaigns that are either from the same actor or a group of actors that use the same infrastructure, independently from the monetizing approach used by the operators of the infrastructure that spreads the samples. 
Analyzing whether profits from a campaign are given to a single actor or a group of actors is out of the scope of this paper.

\descr{Grouping Features}.
We rely on the following features to group samples into campaigns:
    
\descrit{\em Same identifier}:
In order to get rewards from the mining pools, workers must mine using a unique identifier, which in most cases corresponds with the wallet address to which payments are made. In other cases, these are e-mails or other identifiers, like user-generated names. If two samples contain the same identifier, it means that they are accumulating earnings in the same wallet and thus they are grouped together. Some mining tools contain donation wallets to reward the developer, which is done by mining for a certain time (typically 2-5\%) using the donation wallet. While this is configurable and can be turned off, we have observed a few samples doing donations. We note that the CPU cycles donated are also hijacked from the victim and therefore inflict harm to her. 
However we are primarily interested in measuring the earnings of the miscreants, and thus donation addresses are excluded from the aggregation. \review{For this reason, we create a white-list by manually extracting donation wallets from known mining software repositories. We also enrich our white-list using Google searches (e.g., looking for ``Monero'' and ``donation'') and manually analysing the results}.
We have white-listed 14 donation wallets directly obtained from the developers' sites. 
\review{Due to limitations in the manual extraction process, we could be missing donation wallets. 
This can result in the over-aggregation of two independent campaigns as discussed in~\S\ref{sec:discussion}. 
Non--white-listed donation wallets display a characterizing pattern: the same wallet (the donation wallet) appears together with different wallets (from the miscreants) in multiple samples across our dataset. 
However, we do not observe this pattern after white-listing all donation wallets we account for. 
This suggest that we have effectively white-listed all donation wallets.}

\descrit{\em Ancestors}: 
In many cases, the same sample is used to download additional malware. This is the case of droppers, which adapt based on information gathered from the infected host, e.g.: operating system or processor capabilities. 
Accordingly, if a sample is parent of two samples with different wallets, these are grouped together. 
Ancestors and other dropped files that are not directly intended for mining are considered auxiliary binaries and we refer to them as {\tt ancillaries}. This includes samples that do not have a wallet.

\descrit{\em Hosting servers}: 
We use metadata from the samples to extract the URL from where the malware was downloaded. 
A common approach is to host the malware (or even stock mining software) in public cloud storage sites such as Amazon Web Services (AWS), Dropbox or Google Drive (see \S\ref{sec:analysis:landscape}).
Thus, we aggregate two samples if either they are downloaded from the same IP address which does not resolve to a domain from a public repository, or if they are downloaded from exactly the same URL, e.g: \textit{hxxp://suici\-de.\-mouzze.\-had.\-su/gpu\-/amd1.\-exe}. 
We also include the parameters to avoid those cases where a parameter is used to uniquely identify the resource being hosted, e.g.\textit{hxxp://file8desktop.com/download/get56?p=19363}.
This approach has as limitation that we are not aggregating resources where a URL contains ephemeral information (e.g., timestamps or click-IDs), even when they point to the same resource in the server. 
However, this limitation is partially overcome due to other sources for aggregation. 

\descrit{\em Known mining campaigns}: 
As mentioned, we collect IoCs (e.g., domains or wallets) from mining operations reported publicly. 
We look at IoCs that are known to belong to a given mining operation, and look for matches against samples in our dataset. We group two samples if they belong to the same operation.
In our analysis, we have collected IoC for the following mining operations:
Photominer~\cite{Photominer}, Adylkuzz~\cite{Adylkuzz}, Smominru~\cite{Smominru}, Xbooster~\cite{Xbooster}, Jenkins~\cite{Jenkins} and Rocke~\cite{Rocke}. However, our methodology is designed to easily include data collected from new operations.

\descrit{\em Domain aliases (CNAMEs)}: 
During our investigation, we observed many samples using domain aliases (i.e., CNAMEs) that resolve to known mining pools. In these cases, miscreants create one or various subdomains for a domain under their control, and set these subdomains to be aliases of known mining pools. 
Since the resolution is done for the CNAME rather than for the mining pool, they thwart defenses blacklisting mining pools (see~\S\ref{sec:discussion:evasion} for a discussion on anti-analysis techniques). 
To address this evasion method, we perform DNS requests for all the domains extracted from our samples, and look for responses pointing to known mining pools from a CNAME. 
Since CNAMEs might have changed, we also query a DNS history-resolution service provided by AlienVault (\url{https://www.threatcrowd.org}) Accordingly, we aggregate samples using the same domain alias. 

\descrit{\em Mining proxies}: 
Mining using a large number of machines (i.e., more than 100) with the same wallet raises suspicion of botnet usage, and mining pool operators might opt to ban the miner. To prevent this situation, offenders use mining proxies that gather all the shares from the different bots and forward the aggregated to the pool. Thus, pool operators only receive responses from a single machine, the proxy. 
As described in~\S\ref{sec:methodology:analysis}, we identify various samples using proxies. 
We aggregate together samples that use the same proxy. 

\descr{Aggregation}. 
To measure the number of related campaigns and how they are structured, we build a graph where nodes are elements of a given resource (e.g., malware samples, proxies, or wallets) and the edges are determined by the relationships mentioned above. 
We consider each connected component of the graph as a single campaign, where the internal nodes of the graph represent the crypto-mining malware together with the infrastructure used by the campaign. 

\descr{Enrichment}.
After the aggregation, we enrich each campaign with samples related to known Pay Per Install (PPI) services, and mining tools. 
We emphasize that these features are only informative and they are not used to aggregate campaigns. 
We next explain the rationale behind this. 

\descrit{\em Botnets and PPI}:
A common approach to spread malware is through PPI services, where customers pay a fee to botnet operators in order to spread their malware~\cite{Caballero11}. 
Due to commodization of cybercrime services, purchasing a botnet to spread malware is simple and open to anyone with few technical skills, e.g.: by leveraging underground markets~\cite{vanWegberg18}. 
During our analysis, we have observed samples belonging to various botnets that are commonly used as PPI services, such as Virut or Nitol. 
Since these are known third-party infrastructures, two samples using these services are not necessarily related to each other and thus are not aggregated together.

\descrit{\em Stock Mining Software}: 
During our exploratory analysis, we have observed that many campaigns use stock mining software.
This is, the hash of a file dropped by the malware matches with one of the hashes in our collection of mining tools. 
Actually, we have observed that some crypto-mining malware fetch this executable directly from the official GitHub repository. 
However, we have also observed that some miscreants fork these projects and make minor modifications to the mining tool, e.g. to remove donation capabilities.

We use Fuzzy Hashing (FH) to pick up on the aforementioned modifications and to relate these samples with known mining tools. 
FH is a similarity preserving hash function that allows to compare binary files. 
Specifically, FH computes a fingerprint of each binary in such a way that any two binaries that are almost identical map to a ``similar'' hash value. 
Fuzzy hashing has been shown to be an effective way of comparing malware~\cite{li2015experimental}. 
In our pipeline, we use context triggered piecewise hashing~\cite{kornblum2006identifying} and compute the distance between the FH of all samples in a campaign and the FH of all known mining tools. 
We choose a conservative distance threshold of 0.1 as it performs well when comparing malware~\cite{li2015experimental}. 
Thus, samples with a distance lower than 0.1 are considered as stock mining tools. 
\longVer{
\subsection{Summary of data extracted}
Table~\ref{tab:data-samples} summarize the data extracted from each sample. Table~\ref{tab:data-wallets} summarize the data extracted for each XMR wallet and each mining pool.
\begin{table}[]
    \centering
    \begin{tabular}{l|l}
        Name & Description \\
        \hline
        SHA256 & Hash value of the sample \\
        POOL & Normalized name of the mining pool\\
        URLPOOL & URL to which the sample mines\\
        USER & Identifier used to mine in the pool\\
        PASS & Password used to authenticate in the pool\\
        NTHREADS & Number of CPU threads used for mining\\
        AGENT & User agent used for mining\\
        DSTIP & IP to which the sample mines\\
        DSTPORT & Port used for mining\\
        DNSRR & DNS resolutions\\
        SOURCE & Data feeds from which the data was obtained \\
        FS & Date when the sample was first seen\\
        ITW\_URL & URLs hosting or contacted by the sample\\
        PACKER & If any, associated packer used for obfuscation \\
        POSITIVES & Number of positive detections by antivirus* \\
        TYPE & Either \textit{Miner} or \textit{Ancilliary}\\
    \end{tabular}
    \caption{Data extracted for each sample. *Note that the number of AV detecting this malware might increase over time. Thus, the released dataset contain number of positives by April'19.}
    \label{tab:data-samples}
\end{table}

\begin{table}[]
    \centering
    \begin{tabular}{l|l}
        Name & Description \\
        \hline
        POOL & Mining pool\\
        USER & Wallet Identifier\\
        HASHES & Number of hashes shared\\
        HASHRATE & Last hash rate\\
        LAST\_SHARE & Date of the last hash\\
        BALANCE & Total balance (not paid)\\
        TOTAL\_PAID & Total XMR paid\\
        NUM\_PAYMENTS & Number of payments\\
        DATE\_QUERY & Date of last query\\
        USD & Estimated total paid value in USD \\
    \end{tabular}
    \caption{Data extracted for each wallet. 
    In addition to this, for each wallet we also account for the payments and the timestamps extracted from all transparent pools.  Furthermore, for the minexmr pool, we account for the historical hash rate of the wallets.}
    \label{tab:data-wallets}
\end{table}
}
\section{The Binary-based Mining Threat}
\label{sec:analysis}

In this section, we present the analysis of our measurement. 
We first present our dataset (\S\ref{sec:analysis:dataset}), which contains malware seen for over a decade. Next, we perform a longitudinal analysis %
through the lens of our dataset (\S\ref{sec:analysis:landscape}). 
Then, we characterize the type of mining pools and currencies we have seen (\S\ref{sec:analysis:pools}) and study the earnings of the campaigns in the most prevalent cypto-currency, i.e.: Monero (\S\ref{sec:analysis:monero}).

\subsection{Dataset}
\label{sec:analysis:dataset}

Our study results from the collection of 4.5 million malware samples from the range of sources described in~\S\ref{sec:methodology:data}. 
We then apply sanity checks to tailor our analysis to crypto-mining malware only, resulting in a total of: 
\begin{inparaenum}[(i)]
  \item 1,017,110 miner binaries, and
  \item 212,923 ancillary binaries. 
\end{inparaenum}

The samples in (i) are samples where we have observed mining capabilities together with an associated wallet and a pool address. 
The samples in (ii) are samples used by the miners to run the mining operation (e.g., bot clients or loaders). 
In total, our study leverages over 1.2 million crypto-mining malware samples. 
Table~\ref{tab:dataset} shows a summary of our dataset, together with the breakdown of data sources and the type of resources we resort from.\footnote{{See\shortVer{~\cite{extended}} \longVer{Appendix~\ref{sec:appendix-metadata}} for details about other sources crawled.}} 
Our largest source of miners is Virus Total and the smallest is Virus Share. 
As for the resources, we collect the largest number of wallets and pools through dynamic analysis (sandbox and network analysis). The data collection dates to March 2007 to capture the structure of the third party infrastructure from their early stages. However, malicious mining activity starts getting traction in 2011.

\begin{table}
\centering
\begin{tabular}{|c|l|r|}
\hline
\bf Category  & \bf Type & \bf \#Samples \\
\hline
\multirow{3}{*}{Summary} &	ALL EXECUTABLES & 	1,230,033 \\
 & 	 Miner Binaries & 	1,017,110 \\
 & 	 Ancillary Binaries & 	212,923 \\
\hline
\multirow{4}{*}{Sources} &	 Virus Total & 	956,252	 \\
 &  Palo Alto Networks & 	628,915 \\
 & 	 Hybrid Analysis & 	857 \\
 & 	 Virus Share & 	519 \\
\hline
\multirow{3}{*}{Resources} &	Sandbox Analysis & 	1,143,384 \\
 & 	 Network Analysis & 	258,564 \\
 & 	 Binary Analysis & 	10,204 \\
\hline
\end{tabular}
\caption{Our dataset of miners and ancillaries, with the collection sources and the number of resources.}
\label{tab:dataset}
\end{table}

\subsection{Longitudinal Analysis}\label{sec:analysis:landscape}

We extract 16,050 different crypto-mining identifiers from a total of 103,894 samples. 
As mentioned, these mostly include addresses of wallets from various cryptocurrencies, but we also find emails and other identifiers used to authenticate the miner in the pool to later pay them the corresponding reward. 
In the case of wallets, we use regular expressions to detect the associated currency. Regarding the emails, we observe that the majority (97\%) are used as identifiers of one of the most popular mining pools, i.e.: \textit{minergate}.\longVer{\footnote{For a detailed analysis, see\shortVer{~the extended version~\cite{extended}}\longVer{ Table~\ref{tab:pools_emails} from Appendix~\ref{sec:additional-measurements}}.}}

\begin{figure}
\center
\includegraphics[width=.99\columnwidth]{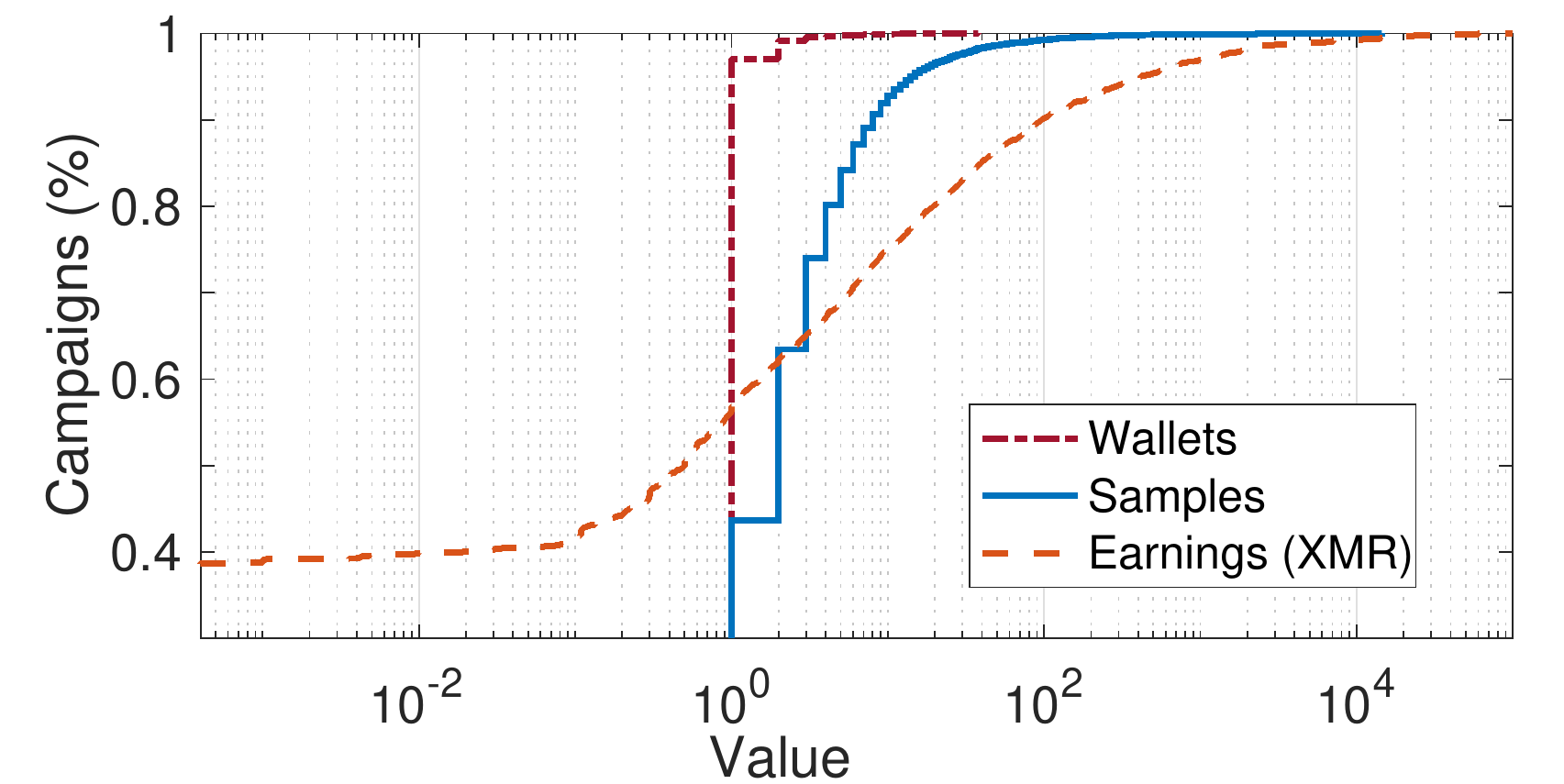} 
\caption{Cumulative distribution number of samples, wallets and earnings observed per campaign.}
\label{fig:cdf-merge}
\end{figure}

Overall, we aggregate samples into 11,387 different campaigns. 
\longVer{Figure~\ref{fig:cdf-merge} depicts the Cumulative Distribution Function (CDF) for the number of samples, and identifiers in all the campaigns. It also shows an overview of the earnings made in pools that provide public statistics for wallets.}
Leftmost side of Table~\ref{tab:currency-campaigns} shows the breakdown of the number of campaigns per type of identifier (i.e., wallets and other identifiers). 
Recall that wallet addresses represent the public key of an electronic wallet in a given cryptocurrency. 
Thus, we show the breakdown for the different cryptocurrencies for which we have wallets. 
Note that two or more identifiers can be used in the same campaign, for example due to a change of a previous wallet address after being banned~\cite{Smominru}.
Monero is the cryptocurrency most frequently used, followed by Bitcoin. \label{finding:underground}
There are at least 18 campaigns using two or more currencies. 
While most of the campaigns are composed by one or few wallets\longVer{ (see Figure~\ref{fig:cdf-merge})}, we observe campaigns having up to 304 different identifiers.

\begin{table}[t]
{
    \centering

    \scalebox{1}{
    \begin{tabular}{|p{5.05cm}|}
    \hline
    \# campaigns with {wallet} addresses for: \\   
    \hline
   \end{tabular}   
    \begin{tabular}{|p{2.6cm}|}
    \hline
     \# samples seen in: \\
    \hline
   \end{tabular}   
   }
   
    \vspace{0cm}

    \scalebox{1}{
    \begin{tabular}{|lr|}
    \hline
Monero     &  2,449  \\
Bitcoin     &  1,535  \\
zCash     &  178  \\
Electroneum     &  150  \\
Ethereum     &  132  \\

\hline
   \end{tabular}
   \begin{tabular}{|lr|}
         \hline
Aeon     &  57  \\         
Sumokoin     &  18  \\
Intensecoin     &  8  \\
Turtlecoin     &  3  \\
Bytecoin     &  1  \\
\hline
   \end{tabular}
    \begin{tabular}{|lrr|}
    \hline
     Year & BTC & XMR \\
     2012 & 9 & 1 \\
     2013 & 23 & 3 \\
     2014 & 223 & 281 \\
     2015 & 115 & 1.6K \\
    \hline
   \end{tabular} 
}

    \scalebox{1}{
    \begin{tabular}{|lp{2.54cm}r|}
    \hline
    Mixed    & &  17\\
    \em Sub-total    &  & 4,548\\ 
    \hline
   \end{tabular}
   \begin{tabular}%
     {|>{\raggedright\arraybackslash}p{.58cm}%
      >{\centering\arraybackslash}p{.58cm}%
      >{\raggedleft\arraybackslash}p{.58cm}|%
    }
     \hline
     2016 & 461 & 8.7K \\
     2017 & 3.8K & 31K \\
     \hline
   \end{tabular} 
    }
    
    \centering
    \vspace{.1cm}

    \scalebox{1}{
    \begin{tabular}{|p{5.05cm}|}
    \hline
    {{ With other identifiers}:} \\   
    \hline
   \end{tabular}   
   \begin{tabular}%
     {|>{\raggedright\arraybackslash}p{.58cm}%
      >{\centering\arraybackslash}p{.58cm}%
      >{\raggedleft\arraybackslash}p{.58cm}|%
    }
    \hline
     2018 & 1.3K & 6.2K \\
    \hline
   \end{tabular}    
    }

    \scalebox{1}{
    \begin{tabular}{|lp{2.47cm}r|}
    \hline
    Email     & &  5,008  \\
    Unknown    & & 2,195  \\    
         \hline
   \end{tabular}
   \begin{tabular}%
     {|>{\raggedright\arraybackslash}p{.58cm}%
      >{\centering\arraybackslash}p{.58cm}%
      >{\raggedleft\arraybackslash}p{.58cm}|%
    }
    \hline
     2019 & 1* & 49* \\
     $\sim$19? & 1.7K & 14K \\
    \hline
   \end{tabular} 
    }
    
    \centering
    \vspace{.1cm}

    \scalebox{1}{
    \centering
    \begin{tabular}{|lp{2.48cm}r|}
    \hline
    \bf TOTAL    &  & 11,751\\ 
    \hline
   \end{tabular} 
   \begin{tabular}%
     {|>{\raggedright\arraybackslash}p{.58cm}%
      >{\centering\arraybackslash}p{.58cm}%
      >{\raggedleft\arraybackslash}p{.58cm}|%
    }
    \hline
     ALL & 7.6K & 62K \\
    \hline
   \end{tabular} 
   }
   
}   
\caption{Leftmost side of the table: number of campaigns per currency, amount of e-mails and unknown identifiers (i.e., not associated with a known currency). Rightmost: number of samples (*partial data) seen in a given year for Bitcoin (BTC) and Monero (XMR).}%
\label{tab:currency-campaigns}
\end{table}

Out of all the samples with wallets, we \review{rely on the {\it first seen} value (obtained from the metadata of the samples provided by Virus Total) to analyse the evolution of number of samples mining Bitcoin and Monero}.  
Rightmost side of  Table~\ref{tab:currency-campaigns} shows the number of samples with wallets by year for the most prevalent cryptocurrencies, i.e.: Monero (XMR) and Bitcoin (BTC). 
Overall, the dataset contains 7.6K BTC and 62K XMR malware samples.\footnote{These are samples with embedded wallets and does not include ancillaries.} 
Judging by the number of samples and the distribution across time, we can confirm %
the decreasing interest in Bitcoin in favour of Monero. Moreover, we have queried available Bitcoin pools with the BTC addresses, and observed negligible earnings (i.e., less than 5K USD).
Note that the data collection ended in early 2019 and thus data from this year is partial. 
Also, due to constraints in the Virus Total rate limit we could not retrieve the {\it first seen} entry for some of the most recently discovered samples.
However, we attribute this samples to 2019 (denoted as {\it $\sim$19?} in Table~\ref{tab:currency-campaigns}). 
\review{As it can be observed, there are 4 samples that were seen in 2012 and in 2013 and that have later been mining Monero, which was released in 2014. This is due to malware reuse, i.e., malware samples that dynamically update their code and execute components downloaded at a later stage (after installation time), namely {\it droppers}. \shortVer{See~\cite{extended} for further details on this.}}
\longVer{For the curious reader, and to foster malware analysis, Table~\ref{tab:xmr-samples-before-2014} show data related to these 4 samples, including hash value and associated wallets. Note that two samples are linked to the same XMR wallet.

\begin{table}[h!]
\centering
\begin{tabular}{l|c|l}
    \multicolumn{1}{c}{MD5} & Year & XMR wallet \\
    \hline
     8b15eb749457b601495c87f465c525f4 & 2012 & 46G5yoqAPP... \\
     2e2ca457803bc6203ddbb5ee4e8855e6 & 2013 & 46G5yoqAPP... \\
     24f08ad6e827f7029141df20f799d6a4 & 2013 & 4BrL51JCc9... \\
     80c30914f32b732868370de6a745bab3 & 2013 & 4JUdGzvrMF... \\ 
\end{tabular}
\caption{Malware samples observed in the wild before 2014 that have later been updated to mine Monero.}\label{tab:xmr-samples-before-2014}
\end{table}

}

\shortVer{We provide more details about the number of samples, wallets, and earnings per campaign in the extended version~\cite{extended}.}%

\longVer{Table~\ref{tab:itw_urls_excerpt} shows an excerpt of the most popular URL domains hosting crypto-miners.}\shortVer{We further look at the most popular URL domains hosting crypto-miners.}\longVer{\footnote{For a complete list ordered by prevalence, see Table~\ref{tab:itw_urls_extended} in Appendix~\ref{sec:additional-measurements}.}} 
We observe that GitHub is the chosen site used to host the crypto-mining malware. 
This is because GitHub hosts most of the mining tools, which are directly downloaded --- for malicious purposes --- by droppers as discussed before. 
Additionally, GitHub is also used to host modified versions of the miners (e.g., by removing the donation capabilities or adding further capabilities). 
It is also used to host ancillary malicious tools~\cite{AvastMalwareGithub}. 
We also observe that there are other public repositories and file sharing sites such as Bitbucket or 4sync, and web hosting sites such as Amazon (AWS), Google, or Dropbox. 
One can also find mining malware hosted through torrent sites (\textit{b-tor.ru}), entertainment sites ({\it telekomtv-internet.ro}), %
or hosted as attachments in the Discord app,  a voice and text chat ({\it cdn.discordapp.com}).
There are also URL-shortener sites ({\it goo.gl}). 
This altogether shows that crypto-miners largely rely on publicly available third-party servers. 
The use of these services provides an economical incentive when compared to other approaches that use dedicated infrastructure such as bullet-proof servers --- that are more resilient against take-downs. 

\vspace{0.5em}
\noindent\fbox{\parbox{\columnwidth}{Our longitudinal analysis confirms previous reports positioning Monero as the preferred currency used by miscreants for crypto-mining malware~\cite{minersPaloAltoNetworks}. 
Thus, in the rest of the paper we focus our attention on campaigns using Monero.}}

\longVer{
\begin{table}[h]
\centering
\begin{tabular}{l|r|r}
\hline
\bf Domains  & \bf \#Samples & \bf \#URLs \\
\hline
github.com &	 163 &	 388\\
*.amazonaws.com &	 85 &	 396\\
www.weebly.com &	 80 &	 96\\
*.google.com &	 38 &	 74\\
hrtests.ru &	 37 &	 1\\
cdn.discordapp.com &	 34 &	 55\\
a.cuntflaps.me &	 32 &	 48\\
file-5.ru &	 30 &	 52\\
\hline
TOTAL \#: 2755(\# domains) &	 3420 &	 6949\\
\hline
\end{tabular}

\caption{ Excerpt of domains hosting known mining malware, number of samples hosted under each domain and number of URLs hosting those samples.}\label{tab:itw_urls_excerpt}
\end{table}
}

\subsection{Mining Pools}
\label{sec:analysis:pools}

There are two possible strategies for mining: joining a pool or mining alone (which we call {\it solo-mining}). 
Using mining pools instead of ``{\it solo-mining} strategies'' has several advantages: it increases the chances of receiving payments for mining and reduces the need for specialized mining equipment. 
Selecting a mining pool is not straightforward because it depends on many dynamic factors such as the current hashrate of the pool, or the complexity required for mining. 
Pools with a high number of workers are more likely to mine a block faster, but the reward received is lower. 
To understand the popularity of the different mining pools among criminals, we look at the number of wallets and the amount of XMR mined over the most consolidated pools (according to various benchmarks such as \url{http://moneropools.com}, or \url{https://minexmr.com/pools.html}) that provide public information about the wallets. 
Table~\ref{table:mining-pools} provides a list of these pools ranked by popularity among criminals (in terms of earnings). 
We show that the most popular pools are \textit{crypto-pool} and \textit{dwarfpool}, with more than 429K and 168K XMR mined respectively.
When looking at the number of wallets observed, the most common pool used is \textit{minexmr}, with (at least) 608 wallets. 
An interesting pool not included in our analysis is \textit{minergate}. 
We have found 4,980 emails mining at this pool in our dataset. 
Since \textit{minergate} does not provide public information about the rewards paid to the miners, we are unable to estimate profits from this pool. 

Our analysis show that 49.3\% of the campaigns use or have used more than one pool.
\longVer{Figure~\ref{fig:num-pools-vs-mined_campaigns} shows the number of pools used by different campaigns grouped by the amount of Monero mined. As it can be observed, the}
\shortVer{Indeed,} 
97\% of the campaigns with largest earnings (i.e., over 1K XMR) have used more than one pool. However, seven campaigns with earnings over 10K are using just one pool. Out of these, six use \textit{dwarfpool} and one uses \textit{crypto-pool}.  
This suggests that mining in different pools depends on different strategies, probably driven by the revenues from each pool and their banning policies.

\longVer{
\begin{figure}
    \centering
    \includegraphics[width=\columnwidth]{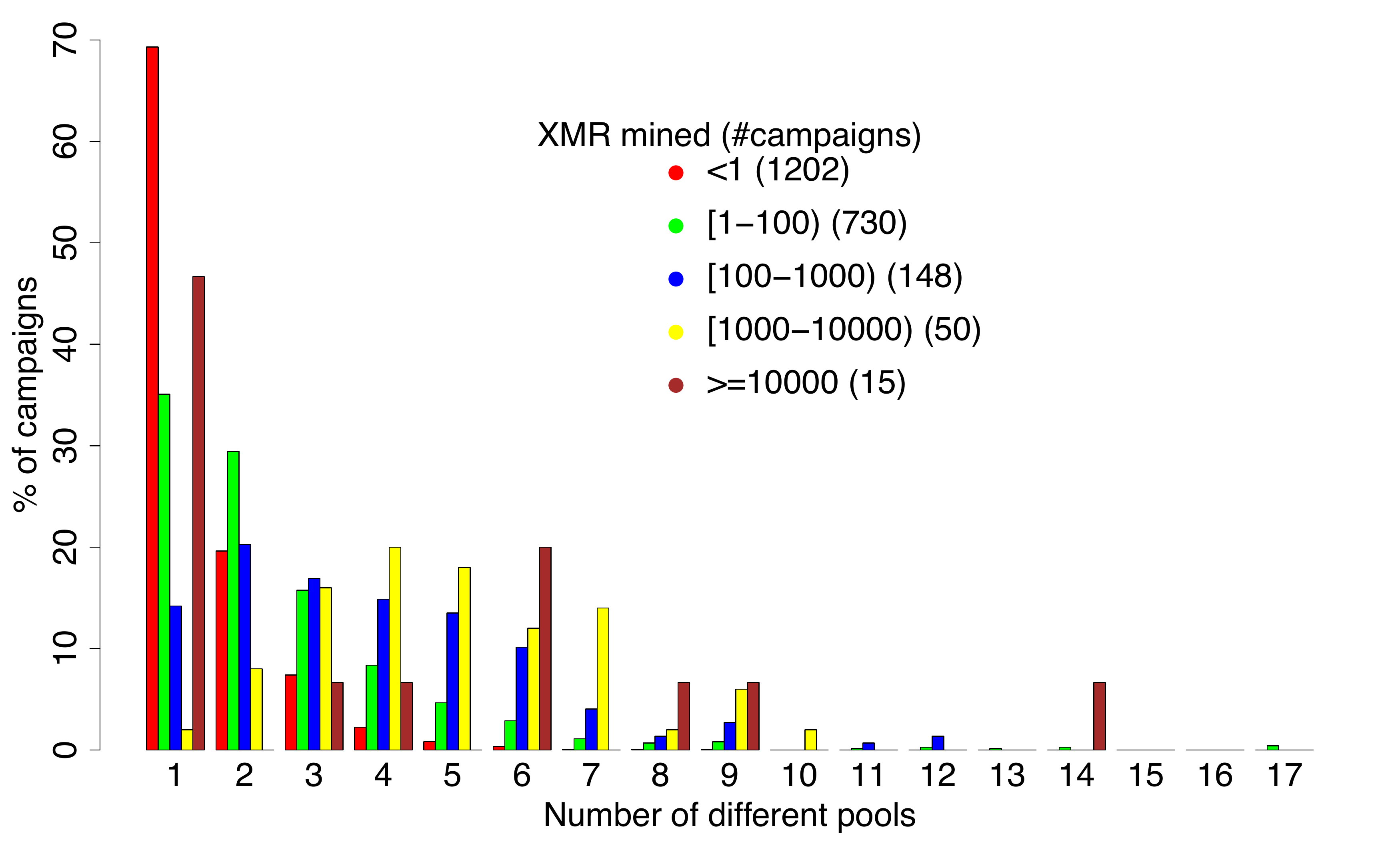}
    \caption{Number of different mining pools used by the campaigns grouped by earnings.}
    \label{fig:num-pools-vs-mined_campaigns}
\end{figure}
}
\begin{table}
\centering
\begin{tabular}{lrrr}
\hline
Pool & XMR Mined &  \#Wallets & USD  \\
\hline
crypto-pool & 429,393 & 487 & 47,261,821\\
dwarfpool & 168,796 & 461 & 1,088,516\\
minexmr & 74,396 & 608 & 5,320,397\\
poolto & 29,044 & 38 & 35,815\\
prohash & 12,833 & 54 & 275,471\\
nanopool & 5,205 & 375 & 858,949\\
monerohash & 4,046 & 217 & 477,557\\
ppxxmr & 3,860 & 185 & 518,487\\
supportxmr & 3,217 & 241 & 443,087\\
Others (8) & 2,797 & 314 & 325,034\\
\hline
\end{tabular}

\longVer{\caption{Overview of the popularity of the different mining pools ranked by the amount of XMR mined by malware.}\label{table:mining-pools}}
\shortVer{\caption{Popularity of the different mining pools ranked by the amount of XMR mined by malware.}\label{table:mining-pools}}
\end{table}

\subsection{Monero-based Campaigns}
\label{sec:analysis:monero}
As shown in Table~\ref{tab:currency-campaigns}, we find 2,360 campaigns mining Monero. %
Out of those, we are able to get payments to 2,145 campaigns through querying the various mining pools.%

We summarize the results of our aggregation in Table~\ref{tab:aggregated_payments} and show some demographics for the top 10 campaigns. Note the difference between USD and XMR in some campaigns. As explained before, this is due to fluctuations of the XMR value and depends on when payments were made. 
A note of precaution when looking at the USD figures as we are unaware of when criminals cash-out their moneros. 
Thus, we prefer to report our findings in primarily in XMR. %
Overall, we estimate that there are at least 2,235 campaigns that have accumulated about 741K XMR (58M USD). Some of them are still active. 
Interestingly, just a single campaign (\FreebufCampaign) has mined more than 163K XMR (20M USD), which accounts for about 22\% of the total estimated. 
This campaign is still active at the time of writing and it is later studied in~\S\longVer{\ref{sec:case:freebuf}}\shortVer{\ref{sec:case-studies}}. 
We observe that only the top 10 campaigns mine more than the remaining 2,225 ones. 
Overall, we observe that 99\% of the campaigns earn less than 100 XMR \longVer{(see Figure~\ref{fig:cdf-merge}).} 
We also observe that while majority of the campaigns earn very little, there are a few campaigns overly profitable. 
This indicates that the core of this illicit business is monopolized by a small number of wealthy actors. 

\begin{table}
\centering
\begin{tabular}{lrrcrr}
\hline
\bf Campaign & \hspace{-1cm} \bf \#S & \bf \#W  & \bf Period & \bf XMR & \bf \$ \\
\hline
C\#627 & 66 & 7 & 06/16 to active* & 163,756 & 20 M\\
C\#3027 & 20 & 2 & 10/16 to 04/18 & 59,620 & 8 M\\
C\#268 & 134 & 4 & 01/15 to 02/19 & 42,069 & 323 K\\
C\#102 & 59 & 1 & 09/14 to 04/18 & 32,886 & 53 K\\
C\#693 & 106 & 2 & 08/14 to 02/19 & 27,985 & 95 K\\
C\#1290 & 91 & 14 & 06/16 to active* & 27,086 & 1 M\\
C\#10465 & 6 & 1 & 09/16 to 04/18 & 23,300 & 2 M\\
C\#3311 & 9 & 1 & 06/16 to 05/18 & 22,520 & 5 M\\
C\#2642 & 46 & 1 & 09/14 to 04/18 & 21,389 & 42 K\\
C\#2202 & 25 & 1 & 09/14 to 04/18 & 20,694 & 38 K\\
\hline
TOP-10 & 562 & 34 & 2014/08/30 - * &441,305 & 38 M\\
ALL-2235 & 64 K & 2532 & 2014/07/18 - * &740,927 & 58 M\\
\hline
\end{tabular}
\caption{Top 10 campaigns ranked by amount of XMR min\-ed. C=Campaign \#S=Num. of samples, \#W=Num. of wallets, and active* on {\tt April 2019}. Recall that the exchange rate to USD is computed dynamically based on when the payments were made. }\label{tab:aggregated_payments}
\end{table}

There are campaigns with a large number of samples, with up to 12K in the case of C\#4 (see further details in our repository).\footnote{\PublicRepo} However, some of the most profitable campaigns have few samples (e.g., C\#10465 or C\#3311).
This means that either the samples in those campaigns have infected a large number of victims, or that other samples from the campaign are not detected by any AV. 
In either case, it suggests that there are some miscreants that are proficient in remaining undetected. In the next section we analyze the infrastructure and stealth techniques used by the different campaigns, and how this affects their efficiency.

While most of profitable campaigns started in 2016 or earlier, we observe recent campaigns with large earnings. 
In particular, 21 campaigns that started in 2018 have mined more than 100 XMR, 12 of which are active at the time of writing (April 2019).

\subsection{Infrastructure}\label{sec:analysis:infrastructure}

We next analyze the third-party infrastructure used in the different Monero campaigns. 

\descr{Mining software.}
\longVer{Table~\ref{tab:tools} shows an overview of the stock mining tools used by the different campaigns.} 
We show that {\tt xmrig}, {\tt claymore} and {\tt niceHash} are the most popular tools we account for. 
With the current distance threshold in our Fuzzy Hashing algorithm, we found no evidence pointing to the use of other less popular tools such as: 
{\tt cast-xmr}, 
{\tt jceMiner}, 
{\tt srbMiner}, or
{\tt yam}.
When using a higher threshold, we found one campaign using {\tt xmr-stak}.
Overall, the top most popular frameworks account for approximately 18\% of the Monero campaigns. 
\longVer{Note that obfuscated versions of these tools are sold in underground marketplaces. 
Thus, these numbers should be viewed as a lower approximation.}

\longVer{
\begin{table}
\centering
\scalebox{1}{
\begin{tabular}{lrlrr}
\hline
\bf Tool & \bf \#I & (\#S)  & \#V & \bf \#C \\
\hline
xmrig   &415    &(299)  &59     &262\\
claymore        &861    &(853)  &14     &98\\
niceHash   &108    &(21)   &11     &67\\
learnMiner      &2      &(2)    &2      &2\\
ccminer &1      &(1)    &1      &1\\
\hline
\end{tabular}
}
\caption{The most popular mining tools used. I=Instances, S=Samples, V=Versions, C=Campaigns}\label{tab:tools}
\end{table}
}

\descr{Domain aliases for mining pools.}
A common mitigation strategy often suggested in commercial reports~\cite{cyberthreatalliance2018} is to block known mining pools, using blacklists. %
Criminals create CNAME domain aliases to evade this mitigation. 
In our analysis, we observe 215 different CNAMEs. 
Most of these are aliases of \textit{minexmr} (176), \textit{cryp\-to-pool} (21) and \textit{dwarf\-pool} (14). 
Interestingly, there are two aliases (\textit{x.\-ali\-buf.\-com} and \textit{xm\-rf.\-fj\-han.\-club}) which have been eventually used to hide two different pools each. 
This suggests again dynamic changes in the mining strategy used by criminals to maximize their revenue. 
We note that the former alias is actually part of the most profitable campaign (\FreebufCampaign), which is detailed in~\S\longVer{\ref{sec:case:freebuf}}\shortVer{\ref{sec:case-studies}}.

\descr{Pay-Per-Install services.}
In order to spread malware, criminals use commodity botnets offered as PPI services in underground markets~\cite{Caballero11,vanWegberg18}. 
We find samples from 3 different botnets offering PPI services. 
In particular, we observe 511 samples associated with the Virut botnet (in 44 different campaigns), 46 from Ramnit (in 10 campaigns) and 27 from Nitol (in 3 campaigns). 
Also, in one of the biggest campaigns (\PhotominerCampaign), known as Photominer~\cite{Photominer}, we find 346 samples (3.01\%) of the samples belonging to this campaign) using Virut to deploy the mining operation.  
Recall that campaigns are automatically extracted. Observing campaigns from botnets that are know to the community shows that our heuristics provide a reliable aggregation. Yet, our framework steps up finding novel campaigns as shown in~\S\ref{sec:case-studies}.

\descr{Obfuscation.}
A common practice when spreading malware is to obfuscate the binary to avoid detection. Criminals typically use existing tools, such as well-known packers (e.g., UPX) or crypters. 
Packers can be fingerprinted more easily than crypters, but crypters --- which are usually purchased in underground markets --- increase the cost of the operation.  
By leveraging the F-Prot unpacker~\cite{F-prot}, we extract packer information associated with each sample (when applicable). \review{This tool also identifies compression algorithms, which are not considered obfuscation. Then, we look at the entropy to detect whether some other unknown obfuscation is applied in samples where no packer or compression algorithm is detected}. 
In our implementation, we choose a conservative threshold of 7.5 (where 8 means total randomness) to decide when a sample is obfuscated, \review{which is more restrictive than values tested in previous works~\cite{mcmillan2011,Ugarte2012}}. 
We found that around 30\% of the samples are obfuscated. 
We consider that a campaign uses obfuscation if a large proportion of their samples (i.e., 80\%) are obfuscated. 
While this is the ratio in the overall dataset, we found that only 4.16\% of Monero campaigns use obfuscation. 
Table~\ref{tab:packers} summarizes the number of samples using obfuscation together with the tool used to obfuscate it. 
UPX is by far the most common tool used. 
Interestingly, we have seen many binaries created using AutoIt (a Windows-based scripting language) which by default packs the script into an PE file using UPX. 
In~\S\ref{sec:discussion:limitations} we discuss the limitations of analyzing obfuscated binaries.

\begin{table}[t]
{
    \centering
    
    \scalebox{1}{%
    \begin{tabular}{|lr|}
    \hline
UPX & 328,493 \\    
NSIS & 17,464 \\
maxorder & 5,988 \\
SFX & 3,928 \\
INNO & 2,423 \\

\hline
   \end{tabular}
   \begin{tabular}{|lr|}
         \hline
eval & 2,032 \\
docwrite & 1,490 \\    
ARJ & 858 \\
CAB & 721 \\
Enigma & 710 \\

\hline
   \end{tabular}
    \centering
    }
    \vspace{.2cm}
    \scalebox{1}{%
    \begin{tabular}{|lp{2.0cm}r|}
    \hline
Others & &4,019 \\
Not packed & & 862,712 \\
         \hline
   \end{tabular}
   }
   
\caption{Packers used for binary obfuscation. %
}
\label{tab:packers}
}
\end{table}

\descr{Analysis.}
Table~\ref{tab:campaigns_analysis} shows the {third-party} infrastructure, stealth techniques and period of activities for the different Monero campaigns (both divided according to their profits, and overall). While only 1.1\% of the campaigns use domain aliases, a higher proportion is found in most profitable campaigns (9.4\% of those mining between 1K and 10K XMR, and 26.7\% of those mining more than 10K XMR). 
A similar situation happens with proxies and PPI services, which are more common in successful campaigns (i.e., with larger earnings). 

Most profitable campaigns have longer period of activity (46.7\% have been active since 2014). However, we also observe a high portion of campaigns (26.7\%) operating only for 1 or 2 years and still having large profits.
We also note the percentage of campaigns active before and after changes in the PoW (Proofs-of-Work): on 06/04/2018, 18/10/2018 and 09/03/2019. These changes require the update of mining software. 
This means that either botnet operators have to update their bots, or customers of PPI services must buy new installs. We show that most of the campaigns stopped due to PoW updates: around 72.4\% in April 2018, 89.3\% in October 2018 and 96.5\% in March 2019.\footnote{Given that our data is from April 2019, some of these campaigns might not be defunct, since it might take some time to update the miners.}
This means that changes in the PoW algorithm might be an effective (though unwitting) countermeasure, as discussed in {\S\ref{sec:discussion:changePoW}}.

\begin{table}
\centering
\begin{tabular}{l@{\hspace{0.1cm}}r@{\hspace{0.2cm}}r@{\hspace{0.2cm}}r@{\hspace{0.2cm}}r@{\hspace{0.2cm}}r}

\hline
    \bf XMR Mined 
  & \multicolumn{1}{c@{\hspace{0.1cm}}}{\bf  $<$ 100}  
  & \multicolumn{1}{c@{\hspace{0.1cm}}}{\bf [100-1k)} 
  & \multicolumn{1}{c@{\hspace{0.1cm}}}{\bf [1k-10k)}  
  & \multicolumn{1}{c@{\hspace{0.1cm}}}{\bf $>$10k} 
  & \multicolumn{1}{c@{\hspace{0.1cm}}}{\bf ALL} \\
\hline
\bf \#Campaigns 
& \multicolumn{1}{c@{\hspace{0.1cm}}}{\bf2,013} 
& \multicolumn{1}{c@{\hspace{0.1cm}}}{\bf154} 
& \multicolumn{1}{c@{\hspace{0.1cm}}}{\bf53} 
& \multicolumn{1}{c@{\hspace{0.1cm}}}{\bf15} 
& \multicolumn{1}{c@{\hspace{0.1cm}}}{\bf2,235} \\
\hline
\multicolumn{6}{c}{\bf THIRD PARTY INFRASTRUCTURE} \\
PPI & 1.3\% & 3.2\% & 9.4\% & 13.3\% & 1.7\% \\
Mining SW & 8.6\% & 14.9\% & 30.2\% & 13.3\% & 9.6\% \\
Both & 0.4\% & 1.3\% & 7.5\% & 0.0\% & 0.7\% \\
\hline
\multicolumn{6}{c}{\bf STEALTH TECHNIQUES} \\
Obfuscation & 4.0\% & 5.2\% & 3.8\% & 0.0\% & 4.0\% \\
CNAMEs & 0.3\% & 5.2\% & 9.4\% & 26.7\% & 1.1\% \\
Proxies & 2.6\% & 6.5\% & 3.8\% & 20.0\% & 3.0\% \\
\hline
\multicolumn{6}{c}{\bf PERIOD OF ACTIVITY} \\
+ Apr-18 & 24.6\% & 57.8\% & 49.1\% & 33.3\% & 27.6\% \\
+ Oct-18 & 9.1\% & 27.3\% & 18.9\% & 33.3\% & 10.7\% \\
+ Mar-19 & 2.7\% & 13.0\% & 5.7\% & 13.3\% & 3.5\% \\
Start: 2014 & 0.2\% & 4.5\% & 11.3\% & 46.7\% & 0.2\% \\
Start: 2015 & 0.2\% & 1.9\% & 3.8\% & 13.3\% & 0.2\% \\
Start: 2016 & 5.5\% & 26.0\% & 41.5\% & 40.0\% & 5.0\% \\
Start: 2017 & 37.3\% & 51.3\% & 41.5\% & 0.0\% & 33.6\% \\
Start: 2018 & 51.7\% & 13.0\% & 1.9\% & 0.0\% & 46.6\% \\
Start: 2019 & 0.5\% & 2.6\% & 0.0\% & 0.0\% & 0.4\% \\
Years: 0 & 69.6\% & 11.0\% & 1.9\% & 0.0\% & 62.7\% \\
Years: 1 & 28.0\% & 57.8\% & 41.5\% & 6.7\% & 25.2\% \\
Years: 2 & 2.2\% & 24.7\% & 39.6\% & 20.0\% & 2.0\% \\
Years: 3 & 0.2\% & 3.2\% & 7.5\% & 20.0\% & 0.2\% \\
Years: 4 & 0.0\% & 3.2\% & 9.4\% & 53.3\% & 0.0\% \\
\hline

\end{tabular}
\caption{Summary of infrastructure, techniques and period of activity for the different campaigns targeting Moneros grouped by profit.}\label{tab:campaigns_analysis}
\end{table}

\label{sec:finding_banned_wallets}

\subsection{Take-Aways}

In summary, the main take-aways of our analysis include:

\begin{asparaenum}

\item We observe that it is no longer profitable to mine Bitcoin, and current criminal efforts focus on mining ASIC-resistant currencies. 
We also show that there are a small number of actors that monopolize the crypto-mining malware ecosystem. 
\longVer{Recent works found similar conclusions in web-based cryptojacking~\cite{Konoth18,Hong18} and crypto-mining malware targeting Bitcoin~\cite{Huang2014} (although this study was from 2014, when mining Bitcoin using desktop computers was profitable).%
}

\item We note that some successful mining campaigns are very complex in terms of the size and infrastructure supporting the campaign. 
Our data shows that about 11\% of these are supported by other underground economies such as third-party  Pay-Per-Install botnets. 
On the contrary, we also observe very profitable mining campaigns that do not appear to use a large supporting infrastructure. 
Instead, they are effective campaigns (due to their long lifetime) with obfuscation and novel evasion techniques, e.g.: using CNAMEs to bypass blacklist-based detection. 

\item We estimate that the malicious ecosystem has currently mined at least 4.37\% of the total Monero in circulation (approximately 58M USD). These numbers must be added to estimations made using web-browser cryptojacking in paralell work.

\item It is common to see campaigns mining in various pools. We observe that the most popular mining pools are {\it crypto-pool}, {\it dwarfpool} and {\it minexmr}. We show that a large number of samples mine to {\it minergate}, an opaque mining pool for which there is no publicly-available information about the rewards received. 

\item When looking at the activity period, we find long-lasting campaigns --- some of which are active at the time of writing. %
In particular, we can see multi-million campaigns operating for a continuous period of more than four years (see Top-10 in Table~\ref{tab:aggregated_payments}). 
This shows that AVs have not addressed this threat appropriately.
We argue that crypto-mining malware has not been given enough attention by the industry and the research community and novel countermeasures are required as discussed in~\S\ref{sec:discussion:countermeasures}. 

\end{asparaenum}

\shortVer{
\section{Case Study}
\label{sec:case-studies}
}
\longVer{
\section{Case Studies}
\label{sec:case-studies}
}

\shortVer{We next present a case study related to a high-profit campaign that has not been previously reported. 
We refer the readers to the extended version of the paper~\cite{extended} for additional case studies.  
Fig.~\ref{fig:case-studies} presents an overview of how the campaign is structured. 
In the graph, nodes in blue represent wallets and nodes in light-green represent malware miners. }
\longVer{We next present two case studies related to high-profit campaigns that have not been previously reported. 
Fig.~\ref{fig:case-studies} presents an overview of how the campaigns are structured. 
In the graphs, nodes in blue represent wallets and nodes in light-green represent malware miners. } 
Thus, various light-green nodes connected to a blue node represent a group of samples using the same wallet. 
Nodes in gray and pink represent the infrastructure of the campaign, with gray nodes portraying contacted domain servers and the pink ones the malware hosts. Finally, ancillary malware are depicted in red and orange. The edges represent the connections described in~\S\ref{sec:methodology:aggregation}.

\longVer{
\begin{figure*}
\centering
\subfloat[The Freebuf campaign (\FreebufCampaign).]{\includegraphics[width=.4\textwidth]{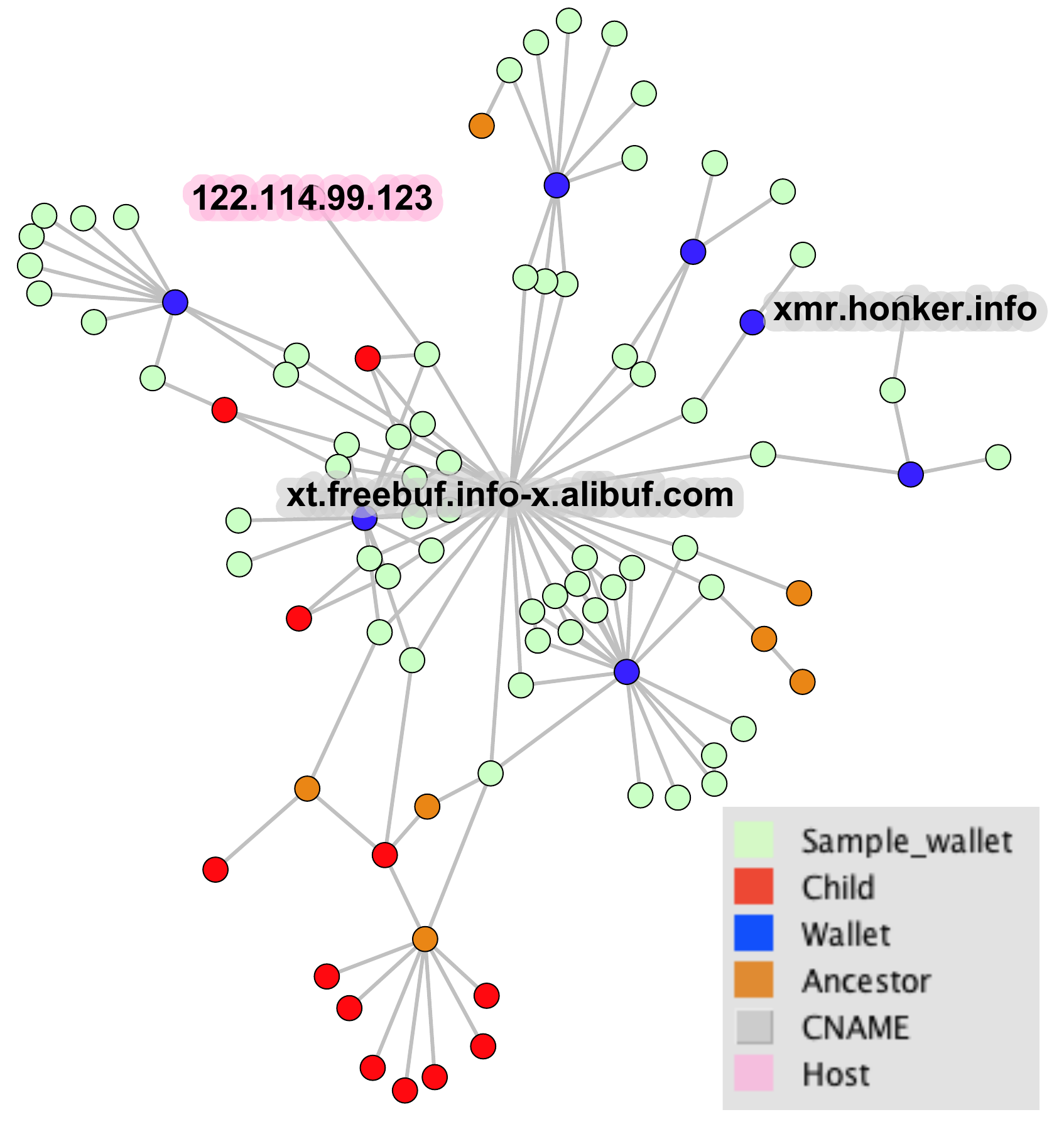}\label{fig:freebuf-graph}}
\hfill
\subfloat[The USA-138 campaign.]{\includegraphics[width=.6\textwidth]{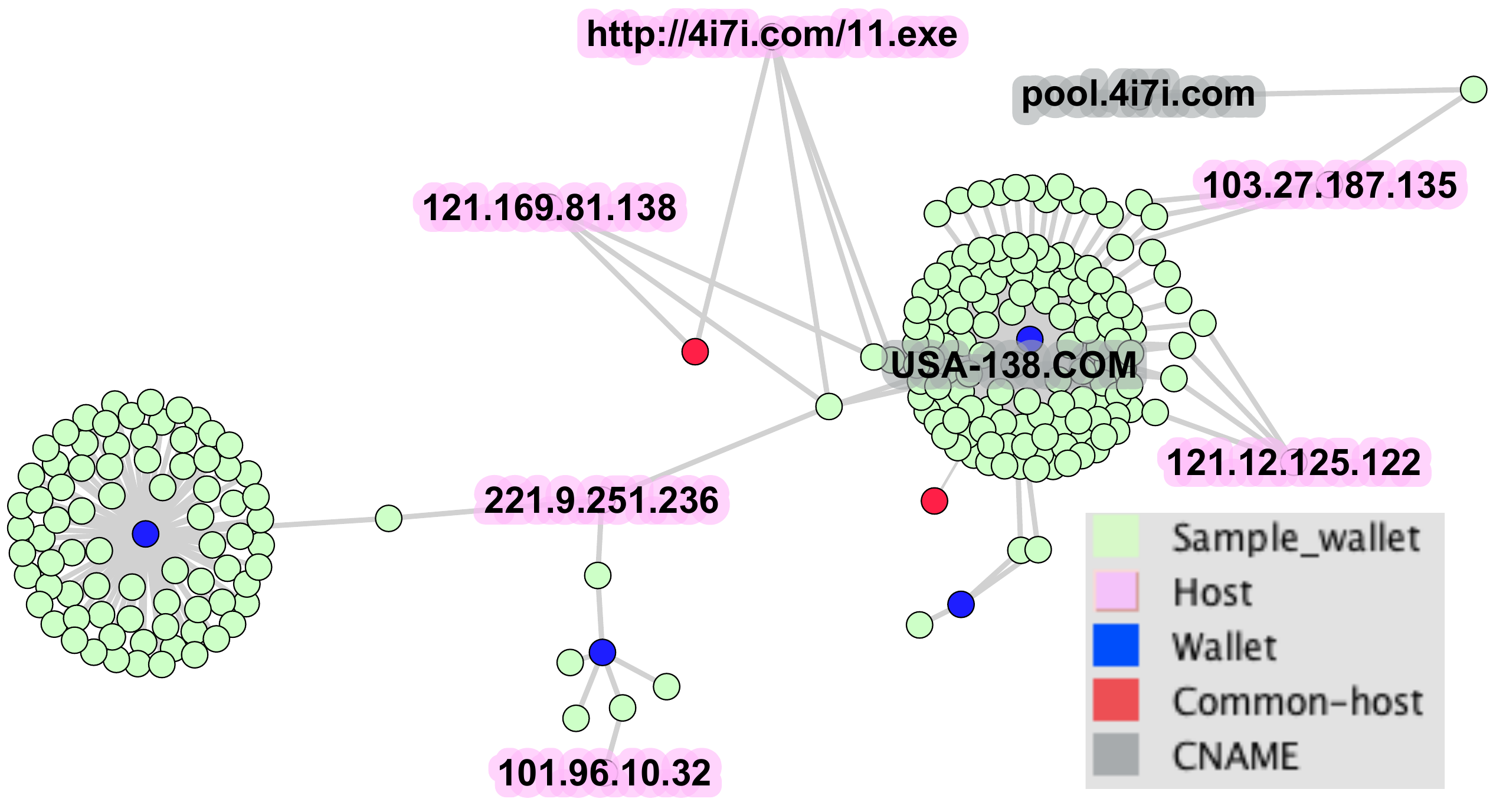}\label{fig:usa138-graph}}
\hfill
\\
\subfloat[Payments per wallet for USA-138.]{\includegraphics[width=\textwidth]{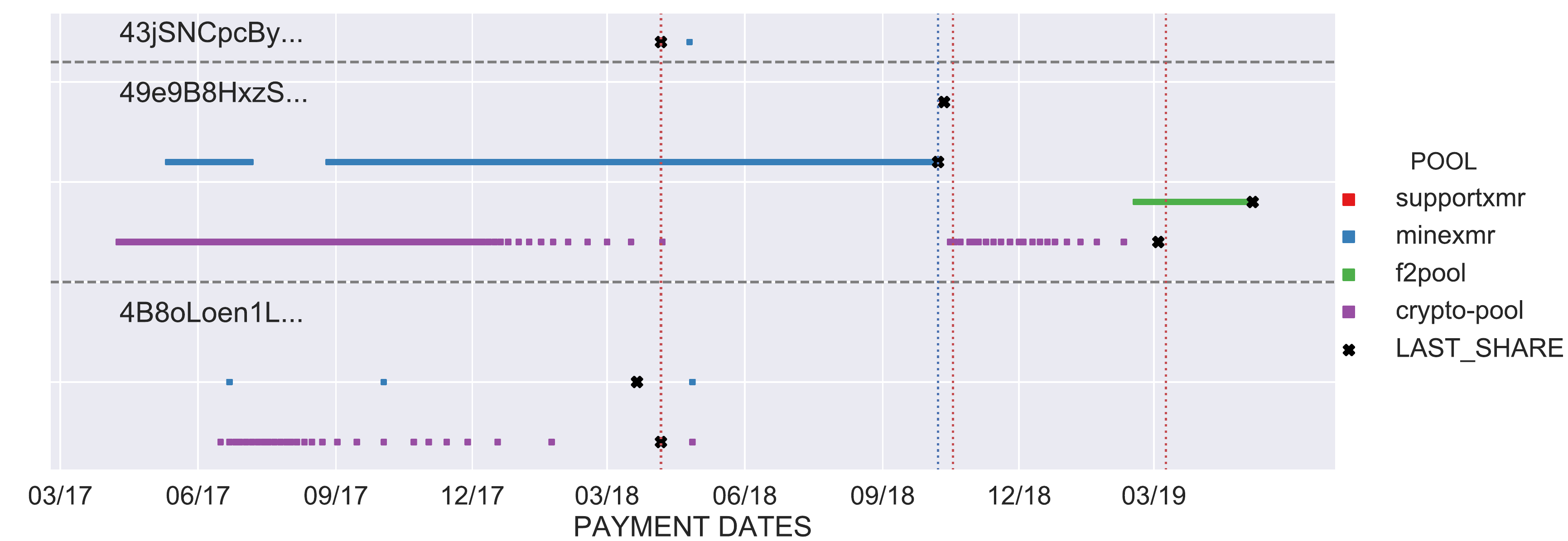}\label{fig:payments-usa138}}
\caption{Overview of the analysis of our case studies. For Fig.~\ref{fig:payments-usa138}, the dashed red lines correspond to the two changes in the Monero PoW algorithm. Dashed blue lines show the day when the wallets were banned in \textit{minexmr}.} 
\label{fig:case-studies}
\end{figure*}
}

\shortVer{
\begin{figure}
\centering
\longVer{\subfloat[The Freebuf campaign (\FreebufCampaign).]}{\includegraphics[width=.4\textwidth]{figures/freebufGraph}\label{fig:freebuf-graph}}
\longVer{\hfill
\subfloat[The USA-138 campaign.]{\includegraphics[width=.37\textwidth]{figures/usa138Graph}\label{fig:usa138-graph}}
\hfill
\subfloat[Payments per wallet for USA-138.]{\includegraphics[width=.36\textwidth]{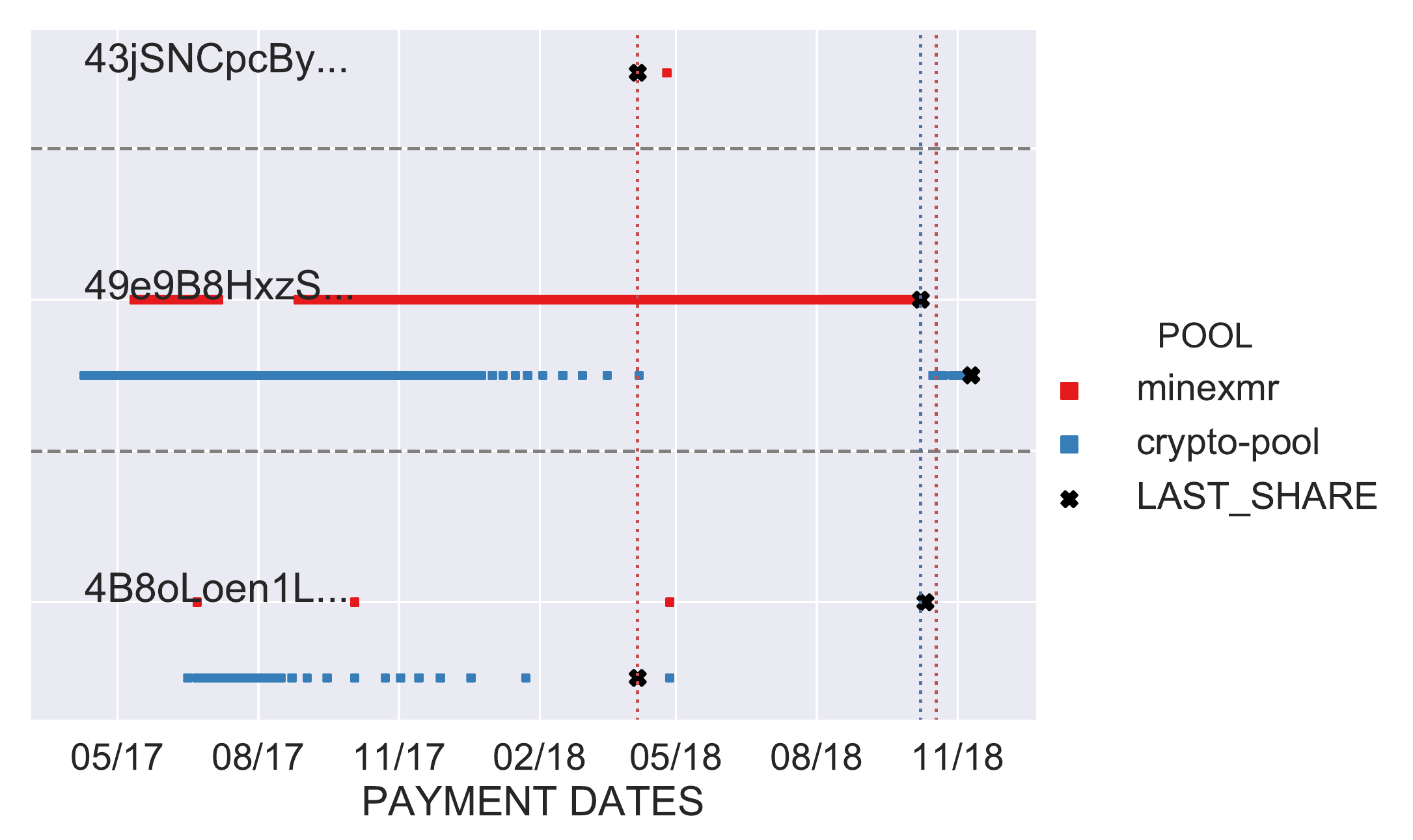}\label{fig:payments-usa138}}}
\caption{Overview of the analysis of our case study. The Freebuf campaign (\FreebufCampaign). 
}
\label{fig:case-studies}
\end{figure}
}

\longVer{\subsection{The Freebuf Campaign}\label{sec:case:freebuf}}

The most profitable Monero campaign (\FreebufCampaign{} in Table~\ref{tab:aggregated_payments}) has mined more than 163K XMR in 3 years using 7 wallets.
We have named it `Freebuf' because the main grouping feature is the domain \textit{xt.freebuf.info}, which is an alias (CNAME) of the \textit{minexmr} pool.

\descr{Structure.}
Figure~\ref{fig:case-studies} shows how the campaign is structured.
We observe that the aggregation is dominated by three grouping features: 
\begin{inparaenum}[(i)]
  \item same identifier, 
  \item ancestors, and
  \item domain aliases (CNAMEs). 
\end{inparaenum}
Interestingly, the graph of this campaign reveals groups of samples with wallets that reach out to one another through paths traversing these three grouping features. 
In other words, the three grouping features are the key to map the structure of the campaign. 
In particular, there are two domains linked through a common wallet: \textit{x.alibuf.com} and \textit{xmr.honker.info}, which in turn are connected to \textit{xt.freebuf.info}. 
Note that both have been aliases of \textit{minexmr}, and \textit{x.alibuf.com} has also been an alias of \textit{crypto-pool}. 
We can observe that the backbone of the graph is established through connections of samples linked to CNAMEs. 
From there, there are other clusters that are linked via {\it same identifier}, and to those, there are some samples which are connected by common ancestors.

\descr{Payments.} 
By analyzing the different payments received through time,
we observe that before the update of the PoW in April 2018, this campaign was mining in various pools simultaneously. 
However, after the update all mining efforts were put into \textit{minexmr}. 
In September 2018 we reported the wallets to the largest pools, resulting in two wallets being banned in October 2018. 
Upon request, one pool operator kindly provided us with statistics regarding the number of different IPs behind the wallets. 
The two banned wallets connected from 5,352 and 8,099 different IPs and had mined 362.6 and 1,283.7 XMR respectively. 
As a consequence of banning, we observe that the campaign operator decided to move their mining efforts to another pool (\textit{ppxmr}) --- which indeed was used before the update on April 2018. We visualize these payments in\shortVer{~\cite{extended}} \longVer{Figures~\ref{fig:payments-freebuf} and~\ref{fig:payments-freebuf-detail} from   Appendix~\ref{sec:case:freebuf-payments}}.
We have seen that as a result of this intervention, together with the change in the PoW algorithm in October 2018, the payments received by the wallets associated to this campaign have been considerably reduced. \newline

\longVer{
\subsection{The USA-138 Campaign}\label{sec:case:usa138}

We have also named this campaign following the name of the CNAME that characterizes this campaign: \textit{xmr.usa-138.com}. 
Overall, this campaign has mined (at least) 7,242 XMR and it has 137 samples and 4 wallets. 
None of the samples from the campaign are known mining software and the campaign does not use proxies. 
Recall that this campaign is still active at the time of writing in December 2018. 
Out of all the samples, 43 are obfuscated with UPX.

\descr{Structure.}
Figure~\ref{fig:usa138-graph} shows how this campaign is structured.
We see that there are two clusters of samples connected through a common host (221.9.251.236). 
This host, which operates in China, is still online at the time of writing and it is still hosting the malware used to run the criminal operation.
One of the wallets (on the left-most side of the graph) is an Electroneum (ETN) wallet (another cryptocurrency based on the CryptoNote algorithm). 
The remaining wallets are Monero. 
We observe a domain, \textit{4i7i.com}, which is used both as a domain alias (\textit{pool.4i7i.com}) for \textit{crypto-pool}, and as a malware host (e.g., \textit{hxxp://4i7i.com/11.exe}). 
We note that the samples using the Electroneum wallet connect to a pool in \textit{etn.4i7i.com}. 
This domain is probably the domain alias of a Electroneum pool, but we are unable to find passive DNS data. 

\descr{Payments.}
As mentioned, the campaign has mined 6,709 XMR {(around 651K USD)}. 
When looking at the amount of earnings made with the Electroneum wallet, we observe earnings of 314,18 ETN. The equivalent of this is currently less than 5 USD. 
While this might seem little at the moment, it might be worth much more in the future. Regarding Monero, this campaign mainly relies on \textit{minexmr} and \textit{crypto-pool}. As depicted in Figure~\ref{fig:payments-usa138}, the most active wallet (named as {\it 49e9B8H...}) operated mainly with \textit{minexmr} after the April 2018 update. 
Once again, we reported the wallets to the pools and soon after we found a similar behavior as with the Freebuf campaign: after the wallet was banned in \textit{minexmr} the malware operators moved to \textit{crypto-pool} again. {Different to Freebuf, this campaign `survived' the PoW change in October 2018 and is still receiving payments from this pool}. 

When looking at the number of connections made by the samples using the wallet {\it 49e9B8H...} to the \textit{minexmr} pool, we observed over 13K IPs. 
These samples have been mining after the update in April'18.
This indicates that the campaign uses a medium-sized botnet which is kept updated. 
}

\section{Discussion}
\label{sec:discussion}

In this section, we analyze existing countermeasures, presenting their main challenges and weaknesses and looking at potential directions to address this challenge. We also discuss the limitations of our work. 
\label{sec:discussion:countermeasures}

\shortVer{
\descr{Reporting illicit wallets.}
\label{sec:discussion:reportingWallets}
Reporting illicit wallets to the pools, while being a common practice~\cite{minersPaloAltoNetworks}, is not an effective countermeasure. First, it is costly and requires cooperation and coordination from all (or at least the main) pool operators. Second, criminals have developed mechanisms to bypass detection (e.g., using mining proxies). During our study, we have reported the illicit wallets we found to the largest pools, together with evidence of criminal behavior. There are non-cooperative pools that chose not to ban wallets; and those that are cooperative pools tend to err on the safe side. For example, the pool \textit{minexmr} only blocks wallets with a large number of associated connections. Since criminals can use proxies to hide botnet-related mining activity, just relying on the number of connections proves ineffective. 

Additionally, we found that successful campaigns mine in several pools. This makes \longVer{campaigns}\shortVer{them} more resilient to take-down operations. Criminals respond to take-downs by changing the mining pool being used (as we have observed in our study) or by creating new wallets and setting proxies up~\cite{Smominru}.
}

\longVer{
\descr{Reporting illicit wallets.}
\label{sec:discussion:reportingWallets}
Reporting illicit wallets to the pools, while being a common --- and important --- practice~\cite{minersPaloAltoNetworks}, is not an effective countermeasure.  First, it is costly and requires cooperation and coordination from all (or at least the main) pool operators. Second, criminals have developed mechanisms to bypass detection (e.g., using mining proxies). 
During our study, we have reported the illicit wallets we found to the largest pools, together with evidence of criminal behavior.
We found two issues. 
First, there are non-cooperative pools that chose not to ban wallets found within crypto-mining malware; and second, those that are cooperative pools tend to err on the safe side. 
For example, the pool \textit{minexmr}, while being remarkably cooperative, has a policy that only blocks wallets with a large number of associated connections. Recall that there are criminals that leverage on a small set of machines (i.e., proxies) to hide botnet-related mining activity. 
Thus, only banning botnet-related mining activity proves ineffective. 

Additionally, we found that many successful campaigns use several pools at the same time. 
While this practice has drawbacks for criminals (mining workers compete with each other), it also makes their campaigns more resilient to take-down operations. 
Criminals respond to such take-downs by changing the mining pool being used (as we have seen in the case studies presented in~\S\ref{sec:case-studies}) or by creating new wallets and setting proxies up~\cite{Smominru}. %

}
\shortVer{
\descr{Changes in the Proof-of-Work algorithm.}
\label{sec:discussion:changePoW}
ASIC-based mining uses customized hardware to compute faster the PoW algorithms~\cite{forkingASIC}. 
Changes in these algorithms are intended to hinder ASIC-based mining, due to the cost of creating new hardware. This also requires updating the software, which is straightforward for benign miners. However, in the case of crypto-mining malware, each change in the algorithm requires botnet operators to update their bots. In turn PPI users have to purchase further installs to push the updated version of their miners. 

We have monitored three changes in the PoW of Monero:  April 6th, 2018, October 18th, 2018 and in March 9th, 2019. 
In each change, around 72\%, 89\% and 96\% of the campaigns ceased their operations respectively. Although changes in the algorithms do not dismantle consolidated campaigns, they can dissuade new ones. Thus, a potential countermeasure against crypto-mining malware is to increment the frequency of such changes, and design these changes to not only be anti-ASIC, but also anti-botnet.}
\longVer{
\descr{Changes in the Proof-of-Work algorithm.}
\label{sec:discussion:changePoW}
ASIC-based mining uses hardware support customized for specific algorithms to compute the faster PoWs~\cite{forkingASIC}. 
Frequent changes in the algorithm are intended to hinder ASIC-based mining, due to the cost of creating new hardware with each change. 
These changes also require updates in the software, which is straightforward for benign miners. However, in the case of crypto-mining malware, it requires botnet operators to update their bots. In turn PPI users have to purchase further installs to push the updated version of their miners. We have monitored three changes in the PoW of Monero:  April 6th, 2018, October 18th, 2018 and in March 9th, 2019.
In each change, about 72\%, 89\% and 96\%  of the campaigns ceased their operations. 
Due to the cost of updating the mining infrastructure, we observe a large number of campaigns not providing valid shares after changes in the mining algorithm (i.e. due to mining with an outdated algorithm). This does not mean that the mining has ceased: a non-updated miner does not provide valid hashes, but it is still mining using infected computers. Thus, the victim is still being harmed as long as the mining continues. 

Although changes in the algorithms do not dismantle consolidated campaigns, they can dissuade new ones. Thus, a potential countermeasure against crypto-mining malware is to increment the frequency of such changes, and design these changes to not only be anti-ASIC, but also anti-botnet.

}
\longVer{
\descr{Security by design and liability.}
In addition to non-co\-ope\-ra\-tive pools, we have observed that a large number of samples mine through opaque pools. 
We position that the community should work on devising protocols that enforce transparency to nodes that act as pools. Mining is a process which requires cooperation from the miners, who get rewards in exchange. Similarly, a potential solution would be to design protocols or techniques that reward transparent and cooperative pools.
Likewise, pools could also improve the mechanisms they use to detect malicious miners, and the actions they take against them. %
Additionally, we argue that a legal framework should be formulated to regulate the pool industry.
}

\label{sec:discussion:limitations}
\shortVer{
\descr{Third-party Infrastructure.}
\label{sec:discussion:third-party-infrastructure}
We observe that most mining campaigns use third-party infrastructure such as PPI services or malware packers. To associate samples to such infrastructure we rely on IoCs gathered from public OSINT repositories, i.e. which have been reported previously. Thus, a limitation of our approach is that campaigns that use such \textit{unknown} third-party infrastructure (e.g. custom packers being offered in underground markets) might be grouped together. Detecting such new third-party infrastructure would require applying other type of intelligence, which is out of the scope of this paper. While this hinders our ability to account for the number of individual actors, our methodology still allows to understand the magnitude of the problem and devising novel mitigation strategies.
}
\longVer{
\descr{Third-party infrastructure.}
\label{sec:discussion:third-party-infrastructure}
We observe that most mining campaigns use third-party infrastructure, both illicit and legitimate. The former includes PPI services, malware packers or private mining pools allowing botnets to mine. The latter includes cloud hosting services such as Dropbox or AWS, and stock mining tools such as \textit{xmrig} or \textit{claymore-tool}. During our analysis, we have observed proficient campaigns making use of both types. %
Using public OSINT, we have analyzed IoCs observed in the samples to associate them to known third-party infrastructure. In particular, we have observed three \textit{known} PPI services (i.e. Virut, Nitol and Ramnit) used by campaign operators to spread their miners across 355K malware variants using \textit{known} packers. However, we have learned about third-party infrastructure (e.g., non-reported botnets, custom malware obfuscators or bullet-proof hosting servers) being offered in underground markets. Thus, a limitation of our approach is that campaigns that use such \textit{unknown} third-party infrastructure (i.e., for which there is no OSINT information) can be grouped together. Detecting such new third-party infrastructure is out of the scope of this study, since this requires investigating each campaign separately and applying other type of intelligence --- by further investigating tools exchanged in underground markets or infiltrating these campaigns. 
Having said it, we have argued that having campaigns grouped by unknown third-party infrastructure, while it hinders our ability to account for the number of individual actors, it is nonetheless useful for law enforcement when prioritizing take-downs and for security practitioners when understanding the magnitude of the problem and devising novel mitigation strategies.
}

\shortVer{
\descr{Quality of the ground-truth.}
\label{sec:discussion:ground-truth}
As with other malware types, a countermeasure to this threat is to use updated AVs. Given the magnitude of the threat, AVs are required to have a large dataset of signatures. However, judging by the activity period of some of the campaigns we have seen that AV vendors have not been up to this task. 
This is a known limitation of the Threat Intelligence industry~\cite{liUsenix2019osint}.

Likewise, our work relies on AV vendors to set our ground-truth. This might introduce both false positives (when legitimate samples are considered malware) and false negatives (when missing actual malware samples). The boundaries between `malicious' and `legitimate' samples are unclear, since a legitimate mining software turns malicious when used in infected computers (i.e., without user consent). Indeed, most AV classify mining tools as malware. We note that establishing an optimal trade-off between benign and malicious mining is not straightforward~\cite{Hong18,Konoth18}. We deal with FP by setting up a relatively high number of positive detections (i.e., 10 AVs). By doing this we are also introducing FN. In this paper, we err on minimizing the number of FP knowing that our findings have to be seen as an under-approximation of the current threat. 
}

\longVer{
\descr{Quality of the ground-truth.}
\label{sec:discussion:ground-truth}
One obvious countermeasure to this threat is to keep educating users to have updated AVs. 
This countermeasure requires AVs  --- provided the magnitude of the threat --- to have a very comprehensive dataset of signatures. 
However, judging by the activity period of some of the campaigns we have seen that AV vendors have not been up to this task. 
This is a known limitation of the Threat Intelligence industry~\cite{liUsenix2019osint}.

Likewise, in our work, we rely on different independent AV vendors to set our ground-truth. 
This might introduce two types of errors that limit our work: False Positives (FP), where a legitimate sample is erroneously flagged as malware, and False Negatives (FN), where malware is flagged as legitimate. Moreover, the boundaries between `malicious' and `legitimate' samples are unclear: a legitimate mining software, when used maliciously in infected computers (i.e. without user consent) might be considered malware (and indeed, most AV classify these tools as malware). We note that establishing an optimal trade-off between benign and malicious mining is not straightforward~\cite{Hong18,Konoth18}.

We deal with FP by setting up a relatively high number of positive detections (i.e., 10 AVs).\footnote{There is one exception to this: we keep a sample with less than 10 AV positives when it contains a wallet observed in another sample having 10 or more AV positives.} 
However, by doing this we are also introducing FN. 
In this paper, we err on minimizing the number of FP knowing that our findings have to be seen as an under-approximation of the current threat. 

We would like to explore a more greedy trade-off by setting the number of positive detections to lower values (e.g., 5 AVs). 
We argue that this should not incur into many FPs as we have introduced a white-listing of known mining tools --- which are more prone to be misclassified. 
In addition, we note that these tools do not contain hard-coded wallets and/or Stratum connections are not observed when run alone in a sandbox.  
Exploring this in detail is precisely the scope of our future work.
}
\review{Finally, it is worth noting that we assume that a wallet is \emph{illicit} when it seen together with a malware. 
However, not all the mining might be illicit. 
For example, the criminal could start mining from his own PC before/while she is mining illicitly. 
This is unlikely, but we would not have technical means to make this distinction should it were to happen. 
While some of the mining could be done licitly, we however argue that the revenues obtained from this activity revert on a criminal anyway. Thus, it can be used to fuel other illicit activities and the figures associated to this wallets are therefore relevant to our study. }

\descr{Quality of the aggregation.}
\review{
Our methodology is used to group samples and wallets into campaigns. 
We leverage the data collected mostly for this purpose, but we also collect additional data to measure the magnitude of the problem. 
We use manual verification to evaluate that our heuristics are coherent.  
When campaigns relate to know mining botnets (e.g., Photominer, or Adylkuzz), we verify that the structure of the campaign matches with what has been reported about that botnet. 
When looking at novel campaigns, we verify that the resulting aggregation matches with the one provided by our heuristics. 
This can be seen from the case studies shown in~\S\ref{sec:case-studies}.
However, we acknowledge that our heuristics are not complete and can be subject to errors in the aggregation. 
This can lead to both campaigns that are under- or over-aggregated. 
In general, we have design our heuristics to be conservative to avoid over-aggregation. 
For example, we aggregate based on the full in-the-wild URL (rather than by domain name) to avoid aggregating campaigns that use common third-party infrastructure like Amazon WS or GitHub. 
}

\shortVer{
\descr{Anti-analysis techniques.}
\label{sec:discussion:evasion}
Our study inherits the limitations of malware analysis. On the one hand, malware uses obfuscation to thwart static analysis, e.g. by using packers and crypters. We have partially addressed this by looking both at the usage of known packers and also at the entropy of the binaries. On the other hand, malware uses evasion techniques to thwart dynamic analysis, (e.g., sandbox detection~\cite{miramirkhani2017spotless}). To partially prevent this, we leverage the use of various sanboxes (e.g. Virus Total and Hybrid Analysis). Also we have attempted to de-anonymize domain aliases that masked connections to mining pools (c.f.,~\S\ref{sec:methodology:aggregation}). 
}
\longVer{
\descr{Anti-analysis techniques.}
\label{sec:discussion:evasion}
A limiting factor of the quality of the ground-truth is the ability of malware to hinder detection. 
On the one hand, malware uses obfuscation to thwart static analysis. 
We have seen that this generally comes in the form of packers and crypters, although advanced adversaries might be using polymorphic or metamorphic malware. 

Like in many of the current countermeasures, we have partially addressed this in two ways.  
First, we looked at the usage of known packers and saw that a small number of campaigns used them. As not all packing algorithms are known, and thus not all samples can be unpacked,  
we have also looked at the entropy of the binaries. 

On the other hand, malware uses evasion techniques to thwart dynamic analysis. 
While there are many forms of evasion we are vulnerable to (e.g., sandbox detection~\cite{miramirkhani2017spotless}), we put special efforts to address those targeting the crypto-mining malware realm in particular.  
Specifically, we have attempted to de-anonymize domain aliases that masked connections to mining pools (c.f.,~\S\ref{sec:methodology:aggregation}). 
However, despite our efforts, our study inherits the limitations of both static dynamic analysis and thus can unavoidably miss samples from advanced adversaries. 
A way to cope with other advanced adversaries that use other forms of evasion, such as anti-emulation techniques, would be to use bare metal solutions~\cite{kirat2014barecloud}. 
}

\shortVer{
\descr{Mining-tailored solutions.}
Miners have a distinctive CPU usage, which allow for anomaly detection based on either modeling the CPU usage~\cite{Konoth18} or on instrumenting web browsers to detect suspicious activity~\cite{Wang18,Kharraz19}. 
However, since the malware controls the infected computer, it might evade local defenses (e.g., by tampering with the CPU monitoring module). Other works propose to monitor the CPU usage of a computer from a hypervisor~\cite{Tahir17}. However, this approach focuses on protecting cloud providers, and  it is not applicable to end-users.
An alternative is to look at the energy consumption fingerprint. While power-aware anomaly detection systems have been proposed in other domains~\cite{kim2008detecting,suarez2015power}, we are not aware of a solution tailored to crypto-mining malware for general-purpose computers. \todo{Ojo a este paper:https://arxiv.org/abs/1909.00268}
We position that these solutions could be deployed by electric-companies to end-users with smart-meters.
}
\longVer{
\descr{Mining-tailored solutions.}
One common strategy when assuming adversaries leveraging advanced obfuscation and evasion techniques, is to devise solutions that are tailored to the type of threat.     
Since miners have a distinctive CPU usage, one can rely on this to build an anomaly detection system for crypto-malware. 
Related works rely on modeling of the CPU usage~\cite{Konoth18} or on instrumenting web browsers to detect suspicious activity~\cite{Wang18,Kharraz19}. 
While these approaches are effective for web-based cryptojacking, these types of defenses are not effective with crypto-mining malware for one reason: the malware controls the infected computer and thus it can evade any local defenses (e.g., by acting as a rootkit and tampering with the CPU monitoring module). Other works propose to monitor the CPU usage of a computer from a hypervisor to protect In\-fras\-truc\-ture-as-a-Ser\-vi\-ce clouds~\cite{Tahir17}. However, this approach focuses on protecting cloud providers, and  it is not applicable to end-users.
An alternative is to offload the usage monitor to an external system and look at the energy consumption fingerprint. 
While power-aware anomaly detection systems have been proposed to detect smartphone malware in general~\cite{kim2008detecting,suarez2015power}, we are not aware of a solution tailored to crypto-mining malware for general-purpose computers. 
We position that these solutions could be deployed by electric-companies to end-users with smart-meters. 
}

\section{Related Work}\label{sect:related-work}

Illicit mining has been a threat since the emergence of Bitcoin in 2009. However, it has not been properly addressed in academia until recently. The first analysis of crypto-mining malware was published in 2014 by Huang et al. \cite{Huang2014}. Authors analyzed botnets and campaigns mining bitcoins. 
\longVer{They found that malicious malware mined at least 4.5K bitcoins (which was worth around \$3.2M in 2014). Since mining bitcoins using end-user computers is no longer profitable, both cryptocurrency malware and web-based cryptojacking rely on cryptocurrencies resistant to ASIC mining, such as Monero or Bytecoin.} 
\shortVer{However, mining non-ASIC resistant currencies (such as Bitcoin) is no longer profitable without dedicated hardware.}
Thus, most of the illicit mining focuses on Monero nowadays~\cite{minersPaloAltoNetworks,Hong18,Konoth18}. 
Recent works analyzed web-based mining, both as an alternative to advertisements to monetize web content~\cite{Papadopoulos18,Ruth18} and as cryptojacking\longVer{, where mining is done without the consent of users}~\cite{Konoth18,Hong18,Saad18}. 
\longVer{Konoth et al. analyzed the Top 1M Alexa sites looking for web-browser cryptojacking ~\cite{Konoth18}. They used a mixture of code analysis and network monitoring to identify whether a web is trying to connect to a mining pool using the Stratum Protocol. Hong et al. proposed a dynamic analysis method to detect cryptojacking in web content~\cite{Hong18}.\newline}
To distinguish cryptojacking from benign mining, it is important to properly identify user consent. One approach is to search for keywords indicating mining activity~\cite{Hong18}. 
This approach misses informed consent acquired by other means, such as images or additional documents. 
Thus, some works also look for AuthedMine scripts, which require explicit action from users to start mining~\cite{Konoth18,Kharraz19}. 
In our work, we rely on AV reports and other heuristics to classify binaries into malware or goodware.

Previous works are characterized by the simplicity in which they aggregate campaigns. 
In particular, related works mostly look at mining pool identifiers (e.g., wallets) alone~\cite{Hong18,Kharraz19}. However, criminals use concurrent miners with different identifiers to retake operations when wallets are banned~\cite{Smominru}. 
Konoth et al. includes information about the servers when performing the aggregation~\cite{Konoth18}.
Unfortunately, this does not scale as it requires manual efforts vetting the code of the scripts (i.e., to get the verification code). 
In their analysis of Bitcoin, Huang et al. use information gathered from the Blockchain to aggregate campaigns~\cite{Huang2014}. 
However, this approach is not valid for cryptocurrencies that obfuscate transactions (e.g., Monero or Zcash).

The overall earnings obtained from malicious mining have increased in the last years. For example, Konoth et al. discovered 1,735 domains, estimating overall revenues of \$188,878 per month. In parallel work, Hong et al. detected 2,770 domains, estimating overall revenues of \$1.7M. 
However, estimations obtained from web-browser cryptojacking are not reliable. 
This is because analyzing profits from web activity relies on estimates of i) the number of monthly visitors, ii) the time spent by each visitor on average, and iii) the type of device they use. 
Instead, we are able to get wallets used by the malware and the payments given by the pools as a reward. 
This allows us to analyze not only the earnings of each wallet, but also the pools used for mining and the exact dates of the payments. 
Our findings increase the understanding of this threat. 
In particular, we estimate that earnings are --- at least --- 58 million USD obtained in 4.5 years of operation (more than 1M/month). 
Table~\ref{tab:related_work} %
summarizes the related works and compares each of the measurements. 

\shortVer{\review{Unlike other type of malware threats, crypto-mining malware is well bounded by a unique characterizing behavior, i.e.: the use of mining activity. 
Thus, simple pattern matching (like the one done by YARA) has proven to be effective.  
As malware increases in sophistication (see discussion in~\S\ref{sec:discussion:limitations}), related works leverage machine learning approaches to systematically identify specific threats~\cite{suarez2017droidsieve}.
However, this is out of the scope of our work. }}

\longVer{
\begin{table*}[bth!]
    \centering
    \resizebox{.9\textwidth}{!}{%
    \begin{tabular}{|l|l|l|l|l|}
    	\hline
         & \multicolumn{1}{c|}{\multirow{2}{*}{Focus (currency)}} %
         &\multicolumn{2}{c|}{Size} & \multicolumn{1}{c|}{\multirow{2}{*}{Profits}}\\
         \cline{3-4}
        & & Analyzed & Detected & \\
        \hline
        Huang et al.~\cite{Huang2014} (2014) & Binary-based mining (BTC) & Unknown & 2K {crypto-mining malware} & 14,979 BTC \\ %
        Ruth et al.~\cite{Ruth18} (2018) & Web-based mining (XMR) & 10M {websites} & 2,287 {websites} & 1,271 XMR/month \\
        Hong et al.~\cite{Hong18} (2018)  & Web-based cryptojacking (XMR) & 548,624 {websites} & 2,270 {websites} & 7,692.30 XMR \\
        Konoth et al.~\cite{Konoth18} (2018)  & Web-based cryptojacking (XMR) &  991,513 {websites} & 1,735 {websites} & 746.55 XMR/month \\
        Papadopoulus et al.~\cite{Papadopoulos18}  (2018) & Web-based mining (XMR) & 3M {websites} & 107.5K {websites} & N/A \\
        Musch et al.~\cite{Musch18}  (2018) & Web-based cryptojacking (XMR) & 1M {websites} & 2.5k {websites} & N/A \\
        \bf Our work & \bf Binary-based mining (various) & \bf 4.4M {malware samples} & \bf 1.2M {crypto-mining malware} & \bf 741K XMR (13.7K XMR/month)\\
        \hline
    \end{tabular}
    }
    \caption{Summary of related work. *Last row shows our measurement.}
    \label{tab:related_work}
\end{table*}
}

\shortVer{
\begin{table}[bt]
    \centering
    \resizebox{\columnwidth}{!}{%
    \begin{tabular}{|l|l|l|l|l|}
    	\hline
         & \multicolumn{1}{c|}{\multirow{2}{*}{Currency}} %
         &\multicolumn{2}{c|}{Size} & \multicolumn{1}{c|}{\multirow{2}{*}{Profits}}\\
         \cline{3-4}
        & & Analyzed & Detected & \\
        \hline
        \cite{Huang2014} (2014) & BTC & Unknown & 2,000 {\malware} & 14,979 BTC \\ %
        \cite{Ruth18} (2018) & XMR & 10M {\textcolor{blue}{\faGlobe}} & 2,287 {\textcolor{blue}{\faGlobe}} & 1,271 XMR/month \\
        \cite{Hong18} (2018)  & XMR & 548,624 {\textcolor{red}{\invertedFaGlobe}} & 2,270 {\textcolor{red}{\invertedFaGlobe}} & 7,692.30 XMR \\
        \cite{Konoth18} (2018)  & XMR &  991,513 {\textcolor{red}{\invertedFaGlobe}} & 1,735 {\textcolor{red}{\invertedFaGlobe}} & 746,55 XMR/month \\
        \cite{Papadopoulos18}  (2018) & XMR & 3M {\textcolor{blue}{\faGlobe}} & 108K {\textcolor{blue}{\faGlobe}} & N/A \\
        \cite{Musch18}  (2018) & XMR & 1M {\textcolor{red}{\invertedFaGlobe}} & 2,500 {\textcolor{red}{\invertedFaGlobe}} & N/A \\
        \bf Our work & \bf Various & \bf 4.4M {\malware} & \bf 1M {\malware} & \bf 741K XMR (13.7K XMR/month)\\
        \hline
    \end{tabular}
    }
    \caption{Summary of related work, where the last row shows our measurement. Legend: \textcolor{blue}{\faGlobe{}} stands for Web-based mining, \textcolor{red}{\invertedFaGlobe{}}  for Web-based cryptojacking, \includegraphics{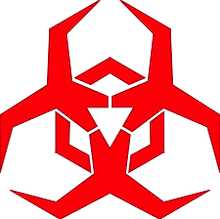} for %
    binary-based crypto-mining malware.
    }
    \label{tab:related_work}
\end{table}
}

\section{Conclusion}

\shortVer{
In this paper, we have presented a {longitudinal} large-scale measurement study of crypto-mining malware --- analyzing the underlying infrastructure relied upon for over a decade. We show that Monero is currently the preferred currency used by criminals, who have obtained massive earnings. 
Our profit analysis on Monero reveals that a small number of actors hold sway the mining illicit business.
We also show that campaigns that use third-party infrastructure (typically rented in underground marketplaces) are more successful. However, this is not always the case. Some of the most profitable campaigns rely on complex infrastructure that also uses general-purpose botnets to run mining operations without using third-party infrastructure. 
Here, we have discovered novel malware campaigns that were previously unknown to the community and we have present technical details about the way this ecosystem operates. 
We have leveraged the insights obtained to discuss countermeasures.
Due to the need of updating mining software, our findings suggest that regular changes in the PoW algorithm discourage criminals.
As future work, we plan to extend our study to emerging campaigns as oppose to looking at the most consolidated ones\review{, and with an extend array of heuristics}.
}

\longVer{
In this paper, we have presented a {longitudinal} large-scale measurement study of crypto-mining malware, analyzing samples spanning over more than a decade. We show that Monero is currently the preferred currency used by criminals, who have obtained massive earnings.
This criminal activity is rooted within the underground ecosystem, which allow criminals to externalize operations, e.g. to avoid AV detection using packers or to spread their malware through PPI.
Through static and dynamic analysis, we extract information from the samples which is used to group them into campaigns. 
While some campaigns rely on {third-party} infrastructure, %
others use simple and effective evasion mechanisms such as {\em domain aliases}.
Our profit analysis on Monero shows that a small number of actors hold sway the mining illicit business. Using crypto-mining malware, criminals have mined (at least) 4.4\% of the moneros in circulation, earning up to 56 million USD. {One of the main reasons of the success of this criminal business is its relatively low cost and high return of investment. Also, since it is considered a lower threat to their clients, the AV industry has not paid due attention.} 
Our findings complement related studies focused on Bitcoin and web-based cryptojacking, corroborating that malicious crypto-mining is a growing and complex threat that requires effective countermeasures and intervention approaches. 
Due to the need of updating mining software, our findings suggest that regular changes in the PoW algorithm might discourage criminals, since this will increase the cost of acquisition (e.g., customers of PPI will need to buy new installs) and maintenance of their botnets.
Finally, we present technical details about the way this ecosystem operates and discuss open challenges and countermeasures. In particular, we analyze two novel campaigns and fully release the data of all campaigns to foster research in the area.
}

\section*{Acknowledgments}

A  shorter  version  of  this  paper  appears  in  the  2019  ACM  InternetMeasurement Conference (IMC). 
This is the full version, but we would like to thank our IMC shepherd, professor Ben Y. Zhao, for the assistance in the process of improving the quality of this paper, and the anonymous reviewers for their positive and helpful comments. 
Special thanks to VirusTotal for granting access to its private API and the Cambridge Cybercrime Centre for sharing the CrimeBB dataset. 
We also thank PaloAlto for sharing data feeds related to cryptomining malware. 
Warmest thanks to Alexander Vetterl and Daniel Thomas for their assistance and valuable feedback.

This work is partially supported by MINECO  (grant TIN2016-79095-C2-2-R), and by the Comunidad de Madrid (P2018/TCS-4566, co-financed by European Structural Funds ESF and FEDER). The opinions, findings, and conclusions or recommendations expressed are those of the authors and do not necessarily reflect those of any of the funders.

\bibliographystyle{IEEEtran}
\bibliography{bibliography.bib}

\appendix

\longVer{
\section{Appendix}
\subsection{Ethical issues}\label{sec:ethical}
}
\shortVer{
\section*{Appendix}
\subsection*{Ethical issues}\label{sec:ethical}
}

Most of the data collected is publicly available. 
However, both Palo Alto and Virus Total shared some non-public information with us and we requested their permission to use it for this paper. 
Another ethical concern relates to the implications of reporting any misuse activity to the pools. 
In particular, we have reported evidence of wallets seen in crypto-mining malware.
Our actions might produce an intervention over the reported users due to criminal activity. 
This entails potential ethical implications when non-criminal wallets are mistakenly banned.
We have taken due precautions to guarantee that we only report wallets of samples used by malware as discussed in the paper. 
Thus, our study has been approved by the designated ethics officer at the Reseach Ethics Board (REB) of our institution.%
\newline
\hspace*{.25cm} In addition to our precautions, we have provided the pools with accompanying evidences proving illicit activity, including a pointer to the Virus Total report. 
However, despite our involvement, the final decision of banning the wallets relies on the pool operators. 
These operators have additional insights about the {\it modus operandi} of their users (e.g., the number of IP addresses that are currently mining with a wallet) that can be used to further corroborate any type of misuse. 
In fact, it is our understanding that the pools that took actions against some of the reported users based their decision solely on the number of connections per wallet. 
Multiple connections from the same wallet evidences the use of a botnet, which it is against the terms and conditions of some of the pools. 
However, we note that a botnet can always hide behind a single IP addressed using proxies. 
Also, we have observed that the pools do not proactively ban wallets that display botnet-like activity. 
We discuss the ethical implications of the banning policies of the pools in~\S\ref{sec:discussion:reportingWallets}.

\longVer{ %
\subsection{Crypto-mining Trends in Underground Forums}
\label{sec:appendix-evolution-threads-forums}
\label{sec:botnet-undergound-forums}

{As discussed in~\S\ref{sec:underground}, the underground economy plays a key role in the proliferation of the malicious crypto-mining threat. 
After analyzing our dataset of posts collected from several underground forums, we observe a large number of posts discussing the use of illicit crypto-malware mining. 
Figure~\ref{fig:evolution-threads-forums} shows the proportion of threats discussing the use of crypto-malware mining across time for different cryptocurrencies. 
We observe that Monero is the most prevalent cryptocurrency in 2018. 
We also observe that while Bitcoin was the most popular cryptocurrency among illicit miners, its popularity has dropped over time. 
We also note that the actors of the underground economy have experimented with other less popular cryptocurrencies such as Dogecoin or Litecoin during the 2013 and 2014. 
However, criminals shifted to Bitcoin and Monero probably when they realized that their value was becoming more profitable.}

{We have studied a large portion of these posts as discussed in~\S\ref{sec:underground}. 
We showed that cybercrime commoditization is key to the wealth of illicit crypto-mining. 
Figure~\ref{fig:botnet-undergound-forums} provides evidence of this.} 
In particular, it shows an advertisement posted in one of the largest English-speaking underground forums, offering a ``Silent XMR'' (i.e. using obfuscated binaries) Botnet. 
Among others, one of the characteristics claimed is the use of xmrig as a mining tool with support for proxies. 
We have seen a wide range of similar adverts in our analysis of underground economies.

\subsection{Other Sources}
\label{sec:appendix-metadata}
{Although our data collection originated from all the data sources described %
in~\S\ref{sec:methodology:data}, we later discovered overlaps between the different sources. 
During our dataset consolidation, we observed that {Virus Total}, {Palo Alto Networks}, {Virus Share}, and {Hybrid Analysis} together accounted for (at least) all the samples observed in the remaining sources. %
We also observed that most of the malicious miners appear in these four datasets.
For this, we consider them to be the main sources for collecting binaries. 
For simplicity, we only refer to {\tt Virus Total}, {\tt Palo Alto Networks}, {\tt Virus Share}, and {\tt Hybrid Analysis} when labeling the source of the dataset in this paper. 
However, we highlight that the alternative sources of data provide valuable complementary metadata that is used in our study. 
We also extend the metadata available for each sample with targeted queries ran using binary or network inspection as described in~\S\ref{sec:methodology:analysis}.}

\subsection{Additional Measurements}
\label{sec:additional-measurements}
In this Section we provide additional measurements obtained from our analysis:
\begin{itemize}

    \item Table~\ref{tab:itw_urls_extended} shows an extension of the URLS hosting the malware observed during our analysis.

    \item Table~\ref{tab:earnings-per-wallet} shows the top 10 wallets sorted by how much they gained. Note that these numbers include donation wallets, which we have filtered for our profit analysis. Nevertheless, we include them in Table~\ref{tab:earnings-per-wallet} to show the similarity between the findings reported in~\cite{minersPaloAltoNetworks}.

    \item Table~\ref{tab:pools_emails} shows the number of emails detected for each associated domain. As it can be observed, most of the emails are used to mine in \textit{minergate}. This is an opaque pool which allows mining in various cryptocurrencies.

\end{itemize}
\begin{table}
\centering
\begin{tabular}{l|r|r}
\hline
\bf Domains  & \bf \#Samples & \bf \#URLs \\
\hline
github.com &	 163 &	 388\\
*.amazonaws.com &	 85 &	 396\\
www.weebly.com &	 80 &	 96\\
*.google.com &	 38 &	 74\\
hrtests.ru &	 37 &	 1\\
cdn.discordapp.com &	 34 &	 55\\
a.cuntflaps.me &	 32 &	 48\\
file-5.ru &	 30 &	 52\\
telekomtv-internet.ro &	 30 &	 30\\
mondoconnx.com &	 26 &	 26\\
free-run.tk &	 25 &	 18\\
brafisaplay1.name &	 25 &	 23\\
b.reich.io &	 23 &	 23\\
mysuperproga.com &	 22 &	 21\\
goo.gl &	 22 &	 32\\
tyme.one &	 21 &	 21\\
gatsoed9.beget.tech &	 20 &	 18\\
mysupflax.name &	 19 &	 18\\
bluefile.biz &	 19 &	 18\\
pack.1e5.com &	 18 &	 16\\
directxex.com &	 18 &	 18\\
dropbox.com &	 17 &	 50\\
*.4sync.com &	 16 &	 142\\
store4.up-00.com &	 16 &	 16\\
www.murphysisters.org &	 16 &	 12\\
fireass.ru &	 16 &	 16\\
weebly.com &	 15 &	 17\\
4.program-iq.com &	 14 &	 15\\
xmr.enjoytopic.tk &	 14 &	 11\\
jkhskdjhsakdjas.info &	 14 &	 7\\
a.pomf.cat &	 14 &	 16\\
giantsto.com &	 14 &	 12\\
daniltinkov228.website &	 13 &	 8\\
root.mcs-katwijk.nl &	 13 &	 8\\
plalium.pw &	 13 &	 12\\
mm.cnxc.tk &	 13 &	 14\\
debittech.ro &	 12 &	 5\\
365experts.com.au &	 12 &	 15\\
www.teamlunyr.com &	 12 &	 12\\
murphysisters.org &	 12 &	 9\\
store6.up-00.com &	 12 &	 13\\
folderfiles10.ru &	 11 &	 20\\
dl.x420.me &	 11 &	 6\\
play.best01011.com &	 11 &	 2\\
garant-ural.ru &	 11 &	 12\\
callfor.info &	 11 &	 17\\
v91049e6.beget.tech &	 11 &	 12\\
\end{tabular}
\caption{ Extended list of domains hosting known mining malware, number of samples hosted under each domain and number of URLs hosting those samples.}\label{tab:itw_urls_extended}
\end{table}

\begin{table}
\begin{tabular}{lrr}
\toprule
Wallet &  XMR mined & USD \\     
\midrule
496ePyKvPB... &  82,985 & 10,655,849 \\
49s5yfpFvE... &  74,643 &  8,964,789 \\
44N9sqiizw... &  55,025 &  7,940,287 \\
454HDLDtqC... &  42,024 &    322,267 \\
42yJMfdGHQ... &  32,886 &     52,830 \\
42NCdZTvv3... &  26,273 &     42,295 \\
44cwDVn9cQ... &  23,300 &  2,288,329 \\
46GGhVFZq8... &  22,520 &  4,775,043 \\
42ychz53ap... &  21,389 &     42,351 \\
46hoCjuFZB... &  20,694 &     37,975 \\
Total for 2,433 wallets & 733,586.75 & 56,605,132.78 \\
\bottomrule
\bottomrule
\end{tabular}
\caption{Amount of Monero mined and corresponding USD for the top 10 wallets.}
\label{tab:earnings-per-wallet}
\end{table}

\begin{table}
\centering
\begin{tabular}{lr}
\bf Pool &\bf \#emails\\
\hline
minergate & 4980 \\
50btc & 41 \\
crypto-pool & 4 \\
supportxmr & 4 \\
nanopool & 4 \\
btcdig & 3 \\
slushpool & 2 \\
moneropool & 2 \\
minemonero & 2 \\
monerominers & 1 \\
monerohash & 1 \\
dwarfpool & 1 \\
suprnova & 1 \\
minexmr & 1 \\
f2pool & 1 \\
OTHERS & 105 \\
\hline
\bf TOTAL & \bf5153\\
\hline
\end{tabular}
\caption{Number of emails in pools}\label{tab:pools_emails}
\end{table}

\subsection{Payments in the Freebuf Campaign}
\label{sec:case:freebuf-payments}

In this section, we visualize the payments done as a mining reward for wallets belonging to the Freebuf campaign, which is the longest campaign and the one with the highest earnings. Concretely, Figure~\ref{fig:payments-freebuf} shows the payments made to all the wallets related to the campaign across time. A more detailed analysis is shown in Figure~\ref{fig:payments-freebuf-detail}, focusing on the payments done in 2018 to the two wallets that were banned in \textit{minexmr}. It can be observed that, after being banned, campaign operators decided to return to the \textit{ppxmr} pool. {However, while the campaign is still active, both the intervention (i.e., wallets being banned) and the change in the PoW of October 2018 has considerably reduced the payments made to this campaign, nearly turning it off}.

\begin{figure}[h]
    \centering
    \includegraphics[width=\columnwidth]{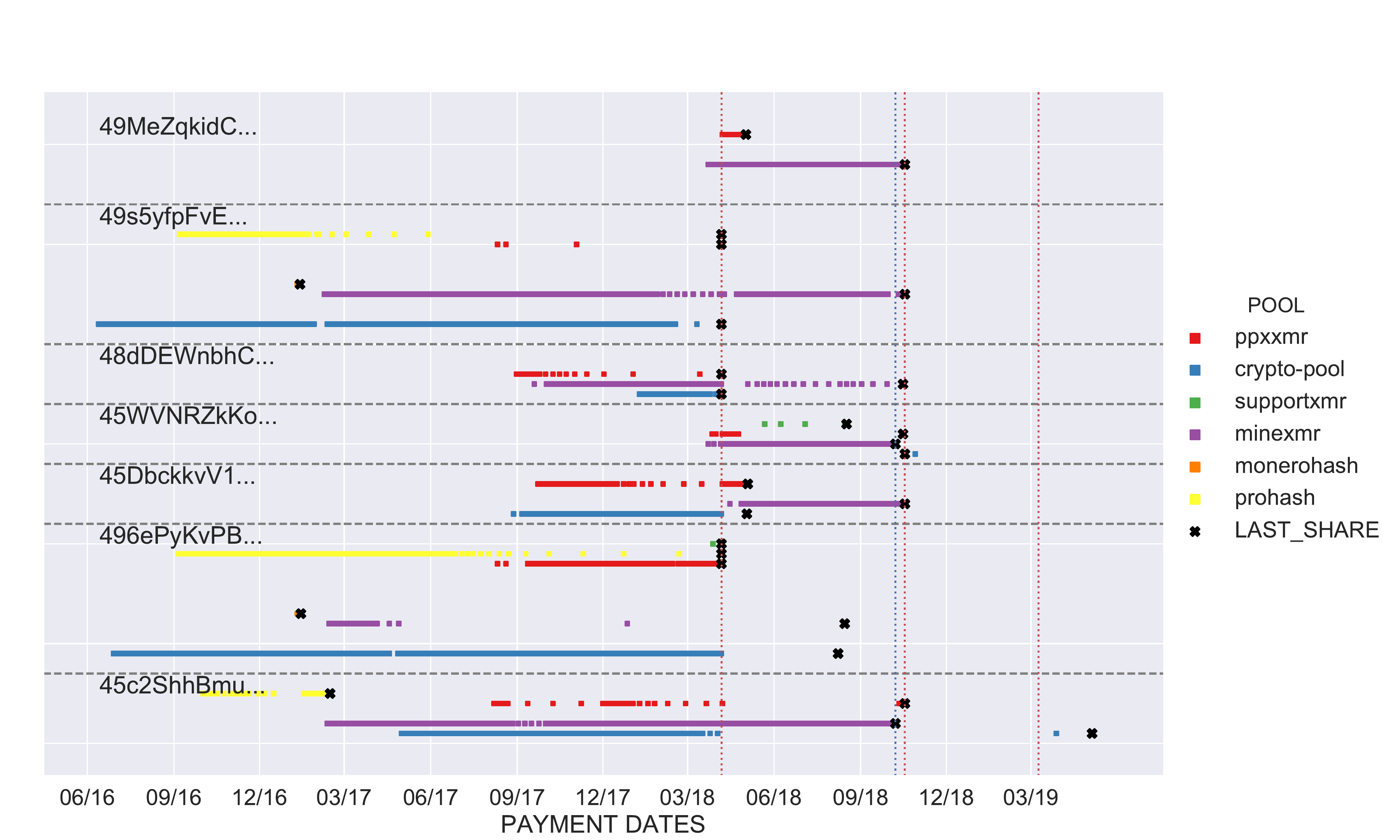}
    \caption{Payments per wallet for the Freebuf campaign. Dashed red lines correspond to the two changes in Monero PoW algorithm.}
    \label{fig:payments-freebuf}
\end{figure}

\begin{figure}[h]
    \centering
    \includegraphics[width=\columnwidth]{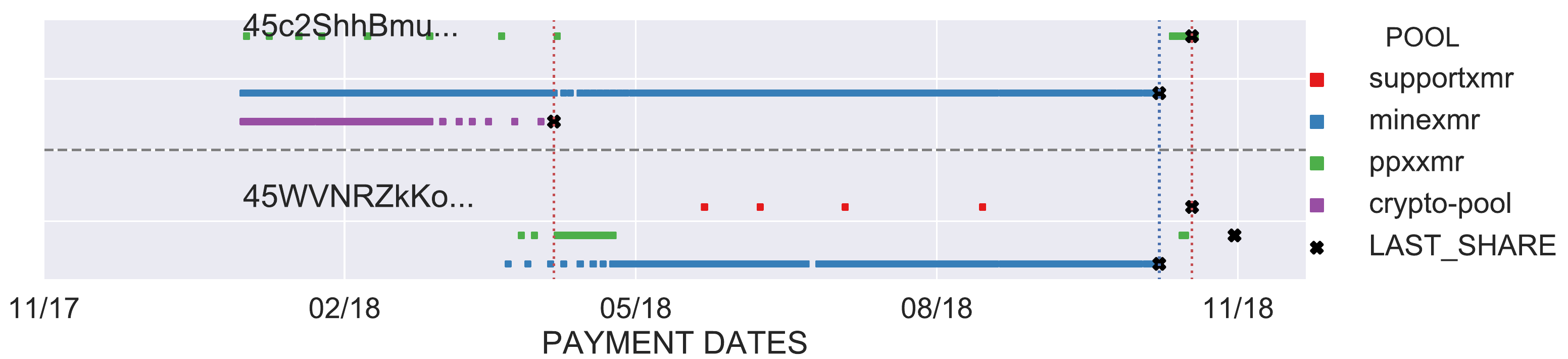}
    \caption{Detailed view of the wallets banned in the Freebuf campaign. Blue vertical line shows when the wallets were banned.}
    \label{fig:payments-freebuf-detail}
\end{figure}

}

\end{document}